\documentclass[%
 reprint,
superscriptaddress,
 amsmath,amssymb,
 aps,
prstab,
]{revtex4-1}

\usepackage{graphicx}
\usepackage{dcolumn}
\usepackage{bm}

\newcommand{\degree}{$^{\circ}$}

\begin{document}

\title{Flux expulsion in niobium superconducting radio-frequency cavities of different purity and essential contributions to the flux sensitivity}

\author{P. Dhakal}
 \email{dhakal@jlab.org}
 \affiliation{Thomas Jefferson National Accelerator Facility, Newport News, VA 23606, USA}
\author{G. Ciovati}
 \affiliation{Thomas Jefferson National Accelerator Facility, Newport News, VA 23606, USA}
 \affiliation{Center for Accelerator Science, Department of Physics, Old Dominion University, \\Norfolk, Virginia 23529, USA}
\author{A. Gurevich}
\affiliation{Center for Accelerator Science, Department of Physics, Old Dominion University, \\Norfolk, Virginia 23529, USA}

\date{\today}

\begin{abstract}
Magnetic flux trapped during the cooldown of superconducting radio-frequency cavities through the transition temperature due to incomplete Meissner state is known to be a significant source of radio-frequency losses. The sensitivity of flux trapping depends on the distribution and the type of defects and impurities which pin vortices, as well as the cooldown dynamics when the cavity transitions from a normal to superconducting state. Here we present the results of measurements of the flux trapping sensitivity on 1.3 GHz elliptical cavities made from large-grain niobium with different purity for different cooldown dynamics and surface treatments. The results show that lower purity material results in a higher fraction of trapped flux and that the trapped flux sensitivity parameter $S$ is significantly affected by surface treatments but without much change in the mean free path $l$. We discuss our results within an overview of published data on the dependencies of $S(l,f)$ on $l$ and frequency $f$ using theoretical models of rf losses of elastic vortex lines driven by weak rf currents in the cases of sparse strong pinning defects and collective pinning by many weak defects.  Our analysis shows how multiscale pinning mechanisms in cavities can result in a maximum in $S(l)$ similar to that observed by the FNAL and Cornell groups and how pinning characteristics can be extracted from the experimental data. Here the main contribution to $S$ come from weak pinning regions at the cavity surface, where dissipative oscillations along trapped vortices perpendicular to the surface propagate into the bulk well beyond the layer of rf screening current. However, the analysis of $S$ as a function of only the mean free path is incomplete since cavity treatments change not only $l$ but pinning characteristics as well. The effect of cavity treatments on pinning is primarily responsible for the change of $S$ without much effect on $l$ observed in this work. It also manifests itself in different magnitudes and peak positions in $S(l)$, and  scatter of the $S$-data coming from the measurements on different cavities which have undergone different treatments affecting both $l$ and pinning. Optimizations of flux pinning to reduce flux sensitivity at low rf fields is discussed.
\end{abstract}

\maketitle


\section{\label{intro}Introduction}

The performance of superconducting radio-frequency (SRF)  cavities is measured in terms of the dependence of the unloaded quality factor $Q_0 = G/R_s$ on the  accelerating gradient, $E_{acc}$, where the factor $G$ depends on the cavity geometry, and $R_s(E_{acc})$ is an average surface resistance.  Recent advances in the processing of bulk niobium cavities have resulted in significant improvements of the quality factor and reducing $R_s$ via diffusion of impurities over a few micrometers from the inner surface of the cavities \cite{Dhakal1,Anna1}. It has been shown both experimentally and theoretically that additional rf losses result from a residual magnetic flux trapped in the superconductor in the form of quantized magnetic vortices during the cavity cooldown through the superconducting transition temperature, $T_c$. Understanding the physics of this process is important to minimize the amount of trapped magnetic flux and reduce the RF losses. For instance, it was found that the amount of trapped flux is affected by the cooling rate, as well as the magnitude and direction of the temperature gradient during the cavity transition to the superconducting state   \cite{HZB1,Roman1, HTC,Fnal1, HZB2,Kubo}.

The typical material used for the fabrication of SRF cavities is bulk, $3-5$ mm thick, fine-grain ($\sim50$ $\mu$m average grain size) niobium with the normal state residual resistivity ratio ($RRR$) of $\sim300$. Large-grain niobium, with grain size typically greater than $\sim1$~cm, is an alternative material for the fabrication of SRF cavities \cite{LG}. One study showed that the losses due to trapped magnetic flux in a large-grain Nb cavity were lower than typically measured in fine-grain cavities of comparable purity and for similar temperature gradients \cite{RG1}. Furthermore, experiments on SRF cavity-grade Nb samples showed that pinning in large-grain Nb is weaker than in fine-grain niobium~\cite{Dhakal2}. The ability to expel flux in fine-grain cavities improved after annealing in a vacuum furnace at 900-1000 \degree C~\cite{PosenTF}, which typically results in grain growth and reduction of density of dislocations.

\begin{figure}[htb]
\includegraphics*[width=70mm]{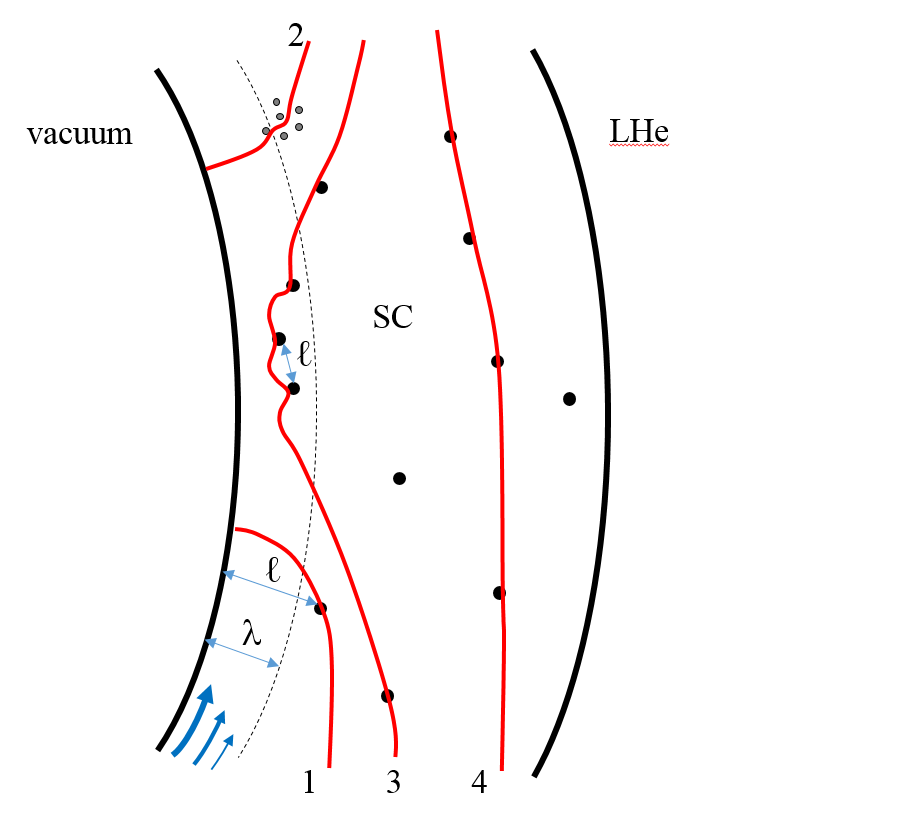}
\caption{\label{fig:pin_sketch} A sketch of the curved cavity wall with trapped vortices in the equator region (not in scale). Dots represent pinning centers, and red lines represent flexible line vortices. The rf current flows in the inner surface layer within the London penetration depth $\sim \lambda$. Vortices 1 and 2 have segments normal to the surface, 1 is pinned by one strong pin, and 2 is pinned collectively by  several weak pins. Vortex 3 has two pinned segments parallel to the surface within $\lambda$. Vortex 4 is not  exposed to the rf field and does not contribute to rf losses.}
\end{figure}

Flux trapping occurs due to pinning of flexible line vortices by materials defects distributed throughout the cavity wall. Yet not all of these vortices contribute to the rf losses as the rf dissipation is due to the oscillation of vortex segments driven by the rf current at the surface. Figure~\ref{fig:pin_sketch} depicts four representative configurations of pinned vortices in the equatorial region of the cavity: normal to the surface, pinned by strong single pins or pinned collectively by array of weak pins, parallel to the surface, or pinned deeper in the bulk.

There can be multiple pinning mechanisms even in high-purity niobium, with stronger pinning by large nonsuperconducting precipitates like hydrides, grain boundaries, dislocation networks and weaker collective pinning of randomly distributed small precipitates or impurities resulting in local variations of mean free path, $\delta l$ or critical temperature, $\delta T_c$ (see, e.g., a review \cite{ce,blatter,ehb}).  It is known that impurities can play a major role in determining the performance of niobium SRF cavities, and treatments such as low-temperature baking (LTB)~\cite{GC_bake} or doping by thermal diffusion~\cite{Dhakal1, Anna1} allow changing the superconducting properties at the surface. Such treatments could not only change the mean free path in the normal state but also affect the spatial distribution, density and strength of pinning centers. Experimentally, the impact of trapped vortices on $Q_0(E_{acc})$ is characterized by a trapped flux sensitivity, $S=R_{res}/B_0$, given by the ratio of the residual surface resistance, $R_{res}$ and the magnitude of the trapped flux, $B_0$. Such quantity reflects the overall dissipation due to vortices trapped by different pinning centers and for different configurations, some of which are shown in Fig.~\ref{fig:pin_sketch}.

Recent studies focused on the dependence of $S$ at low rf field ($\sim 20$~mT) on the mean free path and the frequency \cite{DG1, FnalS, FnalF} of fine-grain, high-purity elliptical cavities. In such studies, different mean free path values resulted from different annealing processes. However, such processes can also alter the pinning characteristics. The objective of this work is twofold:
(i) to evaluate the low-field $S$-parameter in large-grain cavities with different bulk impurities concentration and structural defects to infer the ability of such impurities and defects to pin vortices and (ii) to compare the results with published data and with theoretical models of the rf dissipation of vortices pinned with different orientations with respect to the surface and with different pinning strength.

The paper is organized as follows. In Sec. \ref{Exp_Setup} the experimental setup used for the measurements of $S$ is described.  In Sec. \ref{ExpRes} we present the results of our measurements of the flux sensitivity parameter $S$. In Sec. \ref{sec:comparison} we compare our experimental data with other data published in the literature and fit the data using different theoretical models to infer flux pinning characteristics and other important superconducting parameters. In Sec. \ref{sec:disc} we discuss contributions of different pinning mechanisms to $S$ and the effect of the mean free path on superconducting parameters which control $S$. Sec. \ref{sec:concl} gives the main conclusions of our work.   

\section{\label{Exp_Setup}Experimental Setup}

Three 1.3 GHz single-cell cavities made from discs cut from ingots with different purity were used for this study. The cell shape is that of the cavities for the TESLA/XFEL project~\cite{Tesla},  cavity TC1N1 is a center-cell shape ($G = 269.8$ $\Omega$), cavities G2 and KEK-R5 are end-cell shape ($G = 271.6$ $\Omega$). The cavity name, ingot Nb manufacturer and main interstitial impurities for each ingot are shown in Table~\ref{table1}.

\begin{table*}
\caption{\label{table1}
Purity and manufacturer of the ingots used for the fabrication of the three single-cell cavities used in this study.}
\begin{ruledtabular}
\begin{tabular}{cccccccc}
\textrm{Cavity Name}&
\textrm{Nb ingot supplier}&
\textrm{Bulk RRR}&
\textrm{Ta (wt. ppm)}&
\textrm{H (wt. ppm)}&
\textrm{C (wt. ppm)}&
\textrm{O (wt. ppm)}&
\textrm{N (wt. ppm)}\\
\hline
TC1N1 & Ningxia, China & 60 & $<100$ & 3 & 9 & 100 & 30 \\
KEK-R5 & CBMM, Brazil & 107 & $\sim1034$ & $<10$ & $<30$ & $<30$ & 10\\
G2 & Tokyo-Denkai, Japan & 486 & $\sim81$ & $<0.5$ & $<1$ & $<1$ & $<1$\\
\end{tabular}
\end{ruledtabular}
\end{table*}

The cavity TC1N1 and G2 were fabricated and processed at Jefferson Lab~\cite{cav1, cav2}, whereas the cavity KEK-R5 was fabricated and processed initially at KEK~\cite{cav3}\footnote{The cavity was labeled LG-4 in~\cite{cav3}}. All three cavities were electropolished, removing $\sim 20$~$\mu$m of material from the inner surface, prior to this study.

To explore the effect of the surface preparation on the flux expulsion and the sensitivity of $R_{res}$ to trapped flux, the cavity G2 was re-measured after nitrogen doping. The doping procedure consisted of annealing the cavity at 800 \degree C for 3 hours in vacuum, followed by 2 minutes of exposure to nitrogen at pressure $\sim25$~mTorr. The nitrogen was then evacuated and the cavity temperature was maintained at 800~\degree C for 6 minutes. The cavity was electropolished to remove $\sim 7$~$\mu$m from the inner surface.

Another treatment which affects the near-surface superconducting rf properties of niobium is the LTB. After electropolishing, the cavity KEK-R5 was baked at 120~\degree C for 24 hours in ultra-high vacuum and re-tested.

\begin{figure}[htb]
\includegraphics*[width=80mm]{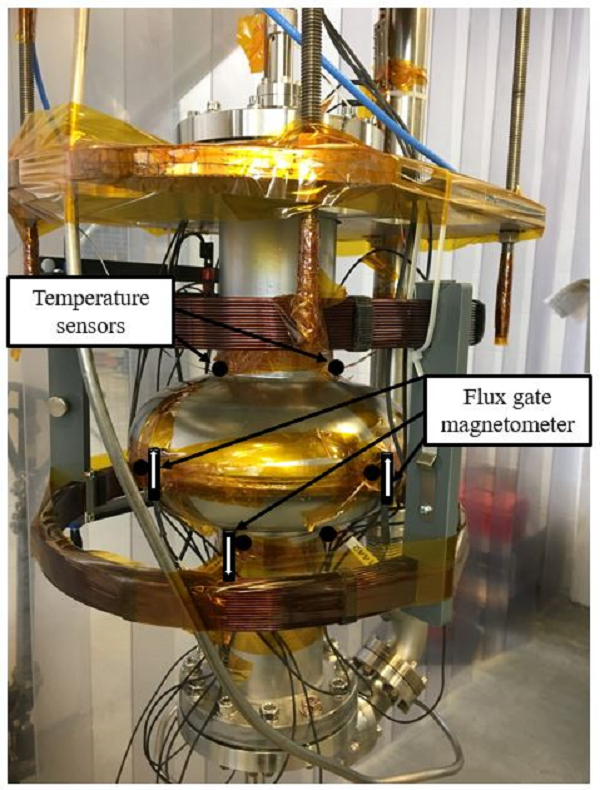}
\caption{\label{fig:setup} Experimental set up of the single-cell cavity with Helmholtz coils, flux-gate magnetometers and Cernox sensors.}
\end{figure}

The setup of the experiment is shown in Fig.~\ref{fig:setup}. 
A Helmholtz coil of diameter $\sim30$~cm was used to create a uniform magnetic field around the cell. Three single-axis cryogenic flux-gate magnetometers (FGM) (Mag-F, Bartington) were mounted on the cavity surface parallel to the cavity axis in order to measure the residual magnetic flux density at the cavity outer surface during the cooldown process. Two magnetic sensors were placed at the equator, $\sim180^{\circ}$ apart, whereas one sensor was placed on the beam tube, close to the iris, to ensure the uniformity of the magnetic flux before the cooldown. The magnetic field uniformity within the cavity enclosure is $\sim \pm1$~mG. Six calibrated temperature sensors (Cernox, Lakeshore) were mounted on the cavity: two at the top iris, $\sim180^{\circ}$ apart, two at the bottom iris, $\sim180^{\circ}$ apart, and two at the equator, close to the flux-gate magnetometers. The distance between the temperature sensors at top and bottom iris is $\sim20$~cm.

The measurement procedure is as follows: (i) the magnetic field was initially set below $2$~mG using the field compensation coil that surrounds the vertical dewar, without any current applied to the Helmholtz coils. (ii) the standard cavity cool-down process was applied, resulting in $\sim4$~K temperature difference between the top and bottom iris, corresponding to a temperature gradient of $\sim0.2$~K/cm. The temperature and magnetic field were recorded until the dewar was full with liquid He and a uniform temperature of 4.3~K was achieved. (iii) $Q_0(T)$ at low rf field (peak surface rf magnetic field $B_p \sim 10$~mT) from $4.3-1.5$~K was measured using the standard phase-lock technique. (iv) The cavity was warmed-up above $T_c$ ($\sim9.2$~K). (v) The cavity was cooled back down to 4.3~K while keeping the temperature difference between two irises below 0.1~K and recording the temperature and magnetic field. (vi) $Q_0(T)$ from $4.3-1.5$~K was measured once more. (vi) The cavity was warmed up above $T_c$ and the current on the Helmholtz coils is set to a certain value. Steps (ii) to (v) were repeated for three different values of magnetic field.

\begin{figure*}[htb]
\includegraphics*[width=150mm]{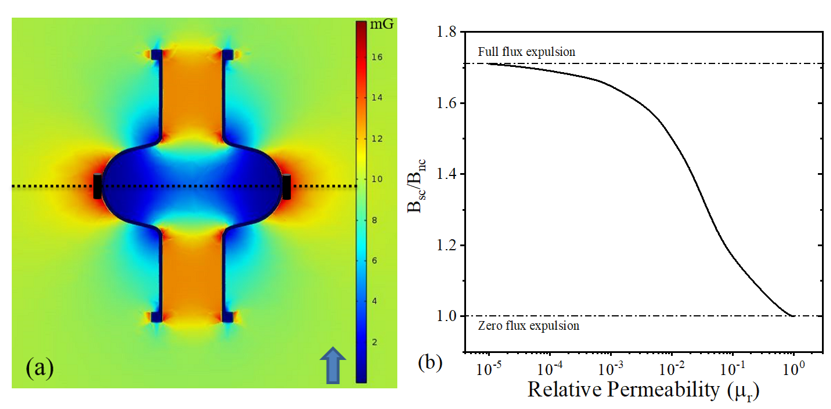}
\caption{\label{fig:ComSim} (a) Contour plot of the magnetic field distribution around the perfectly diamagnetic cavity, $B_{sc}$, with an applied axial uniform magnetic field $B_n=10$~mG, shown by the arrow. (b) The flux expulsion ratio as a function of relative permeability ($\mu_r$) of the bulk Nb at the center of the FGM at the equator.}
\end{figure*}

Fig.~\ref{fig:ComSim} shows the results of a magnetostatic finite element analysis using the software COMSOL~\cite{comsol} for a single-cell cavity of the same geometry as the one used for our experiments.
A magnetic field of 10~mG was applied along the cavity axis and the color map shows the distribution of the magnetic field calculated for a perfectly diamagnetic cavity in the ideal superconducting state. Figure~\ref{fig:ComSim}(b) shows the ratio of the magnetic field just outside the equator in the superconducting state divided by the applied field as a function of the permeability of the cavity. Different values of permeability represent different amount of trapped magnetic field.

\section{\label{ExpRes}Experimental Results}

\subsection{\label{sec:cooldown}Cool-down and flux expulsion}

\begin{figure}[htb]
\includegraphics*[width=80mm]{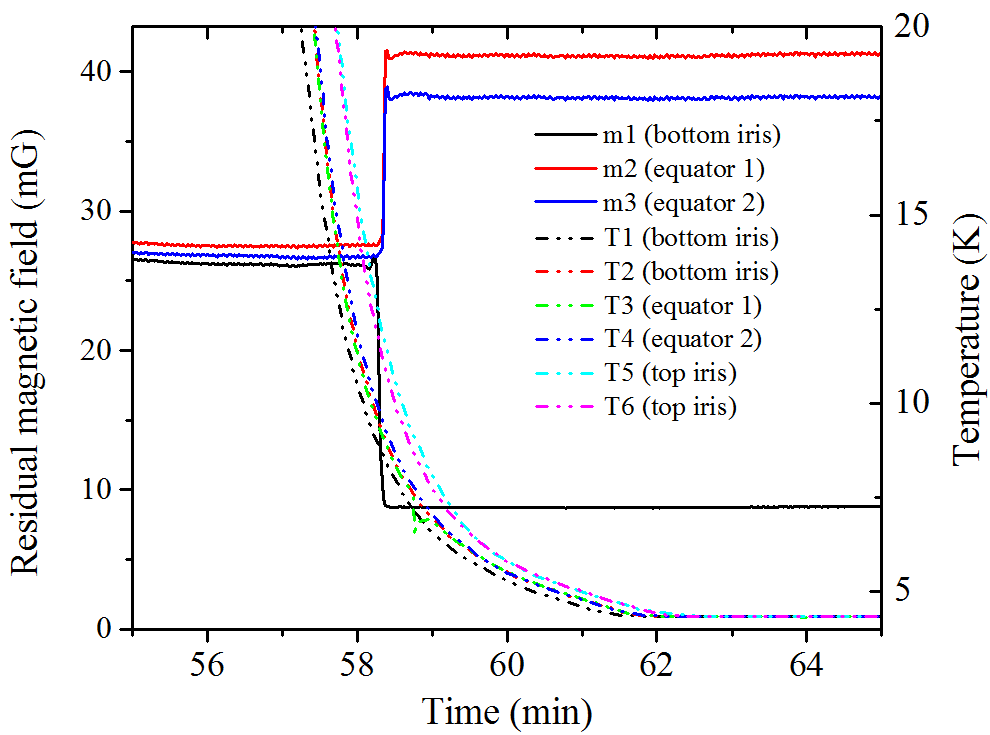}
\caption{\label{fig:BnT} Temperature and magnetic field during transition from normal to superconducting state measured during a cool-down cycle of cavity G2.}
\end{figure}

The ratio of the residual dc magnetic field measured after ($B_{sc}$) and before ($B_n$) the superconducting transition qualitatively explains the effectiveness of the flux expulsion during the transition. A value of $B_{sc}/B_n = 1$ represents complete trapping of magnetic field during cooldown, whereas a flux expulsion ratio of 1.7 at the equator and 0.4 at the iris would result from the ideal superconducting state, as shown in Fig.~\ref{fig:ComSim}(b). Experimentally, $B_{sc}/B_n$ depends on the Nb material and on the temperature gradient along the cavity axis during the cool-down. Values of $B_{sc}/B_n$ close to the theoretical estimate could be achieved with high temperature gradient ($\Delta T > 10$~K)~\cite{Roman1, PosenTF, RG1, DG1}.
A representative plot of the residual magnetic field at the FGMs locations measured during one cool-down cycle for cavity G2 is shown in Fig.~\ref{fig:BnT}. The average value of $B_{sc}/B_n$ for the two FGMs at the equator was $1.45 \pm 0.05$, whereas $B_{sc}/B_n = 0.35$ for the FGM close to the iris. The jumps in magnetic flux density occurred at 8.9~K for sensor m1, 9.1~K for sensors m2 and 9.3~K for sensor m3. The temperature difference between the top and bottom iris when the bottom iris reached 9.2~K was 2.6~K.

\begin{figure}[htb]
\includegraphics*[width=80mm]{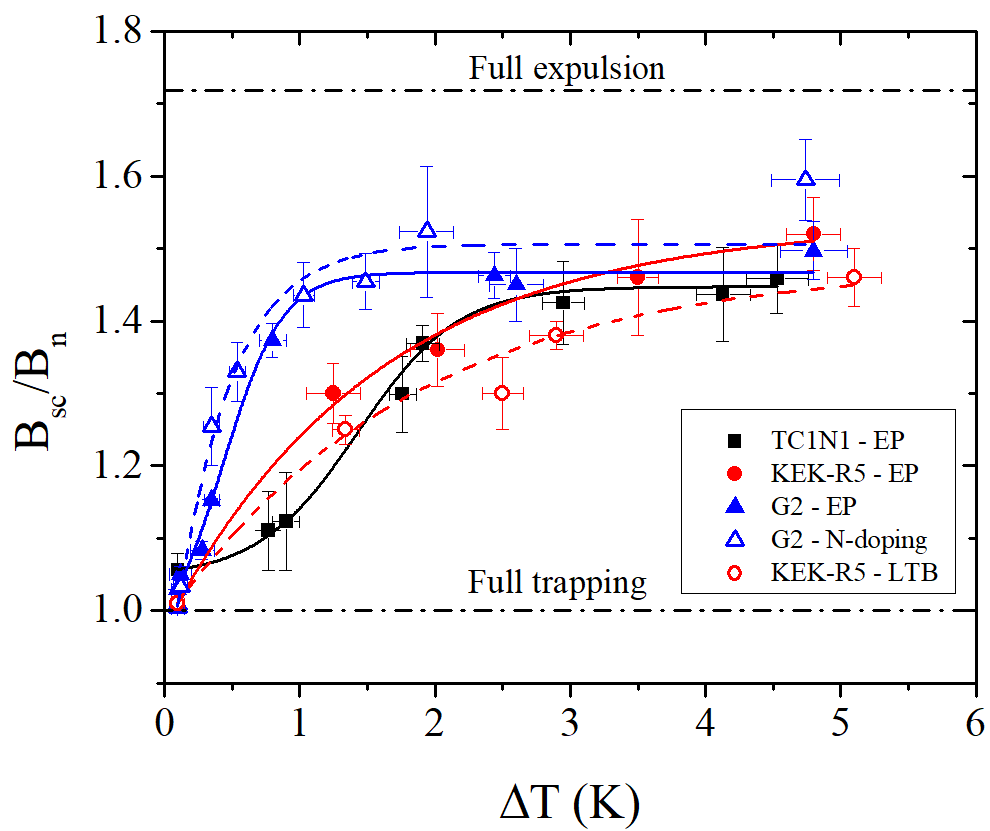}
\caption{\label{fig:FluxExp1} Average flux expulsion ratio at the equator as a function of the temperature difference (iris-to-iris) on cavities after EP surface treatment, N-doping (G2) and LTB (KEK-R5). The lines are sigmoidal fits to the data and are a guide to the eye.}
\end{figure}

Figure~\ref{fig:FluxExp1} shows the average flux expulsion ratio at the equator measured for the three cavities (TC1N1, KEK-R5 and G2) after removal of $\sim20$~$\mu$m from the inner surface by electropolishing and after N-doping of cavity G2 and LTB of cavity KEK-R5. All three cavities showed good flux expulsion with $B_{sc}/B_n \sim 1.5$ when the temperature difference between irises was greater than 4~K.

\subsection{\label{sec:rfdata}rf measurements}

\begin{figure}[htb]
\includegraphics*[width=80mm]{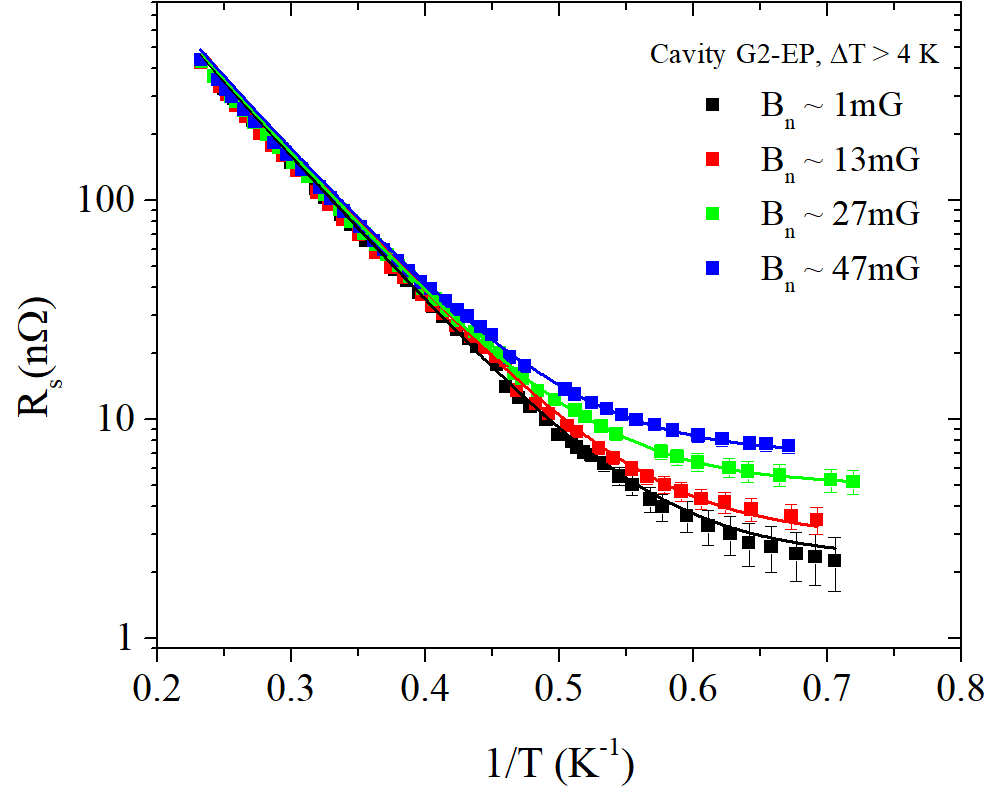}
\caption{\label{fig:RsT} $R_s(T)$ measured in electropolished cavity G2 for cool-downs with $\Delta T > 4$~K with different applied dc magnetic field values prior to cool-down. Solid lines are fits with Eq.~(\ref{eq1}).}
\end{figure}

The average rf surface resistance was obtained from the measurement of $Q_0(T)$ at low rf field ($B_p \sim 10$~ mT) for two different cool-down conditions, one with uniform temperature ($\Delta T < 0.1$~K) and one with high temperature gradient ($\Delta T > 4$~K). Such measurements were repeated at different applied dc magnetic field, $B_a$, prior to each cool-down. The $R_s(T)$ data are shown, as an example, in Fig.~\ref{fig:RsT} for cavity G2. The data were fitted with the following equation:
\begin{equation}
R_s(T) = R_{BCS}(T,\omega,l,\Delta) + R_{res},
\label{eq1}
\end{equation}
where the BCS surface resistance $R_{BCS}$ was computed numerically from the Mattis-Bardeen (M-B) theory~\cite{MB} using the Halbritter code~\cite{Halb}. The mean free path, $l$ and the ratio $\Delta/k_BT_c$ were regarded as fit parameters, where $\Delta$ is the energy gap at $T=0$, and $k_B$ is the Boltzmann constant. We took $T_c = 9.2$~K, the coherence length, $\xi _0 = 39$~nm and the London penetration depth, $\lambda_0 = 32$~nm for Nb in the clean limit, $\xi_0\ll l$ at $T=0$.

The values of $l$ and $\Delta /k_BT_c$ did not change, within experimental uncertainty, with different cool-down conditions or applied dc magnetic field $< 50$~mG. The weighted average values of $l$ and $\Delta /k_BT_c$ from eight data sets for each cavity are shown in Table~\ref{table2}. These mean free path values indicate that the surfaces of all three cavities were in a moderate dirty limit $l\lesssim \xi_0$. The extracted value of $\Delta/k_BT_c$ is $\sim20 \%$ lower in the low-purity cavity as compared to the other two. Since the mean free path may vary over the scale $\lesssim \lambda(T)$ perpendicular to the surface the temperature range used to extract $l$ is indicated between parenthesis in Table~\ref{table2}.

\begin{figure}[htb]
\includegraphics*[width=100mm]{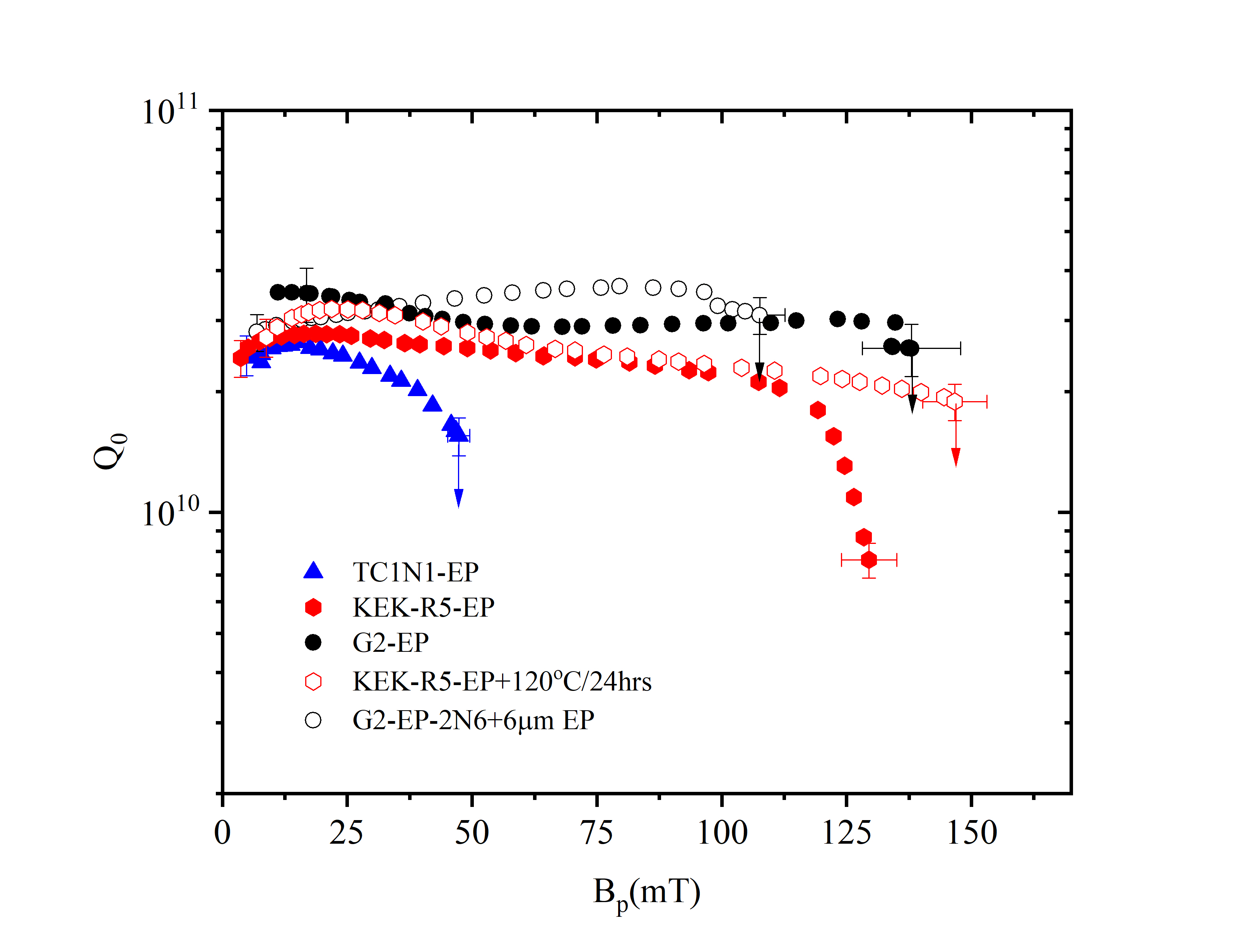}
\caption{\label{fig:QvsE} $Q_0(B_p)$ measured at 2.0~K after cool-down with $\Delta T > 4$~K and $B_n < 5$~mG for each cavity and treatment listed in Table~\ref{table2}. The arrows indicate the quench field.}
\end{figure}

The curves of $Q_0(B_p)$ measured at 2.0~K after cool-down with $\Delta T> 4$~K and $B_n < 5$~mG for each cavity and treatment listed in Table~\ref{table2} are shown in Fig.~\ref{fig:QvsE} and they are fairly typical for those treatments. There was no field emission in any of the tests. A multipacting barrier occurred at 136 mT during the test of G2, causing a drop in the $Q$-value.

\begin{table*}
\caption{\label{table2}
$S$, $R_{res0}$ and fraction of the applied field being trapped, $\eta_t$, obtained from fits of $R_{res}(B_n)$ for different cool-down conditions and weighted average values of mean free path and $\Delta /k_BT_c$ obtained from fits of eight data sets of $R_s(T)$ between $1.5 - 4.3$~K for each cavity.}
\begin{ruledtabular}
\begin{tabular}{cccccccc}
\textrm{Cavity Name}&
\textrm{Bulk RRR}&
\textrm{Treatment}&
\textrm{$l (1.5-4.3$~K) (nm)}&
\textrm{$\Delta /k_BT_c$}&
\textrm{$R_{res0}$ (n$\Omega$)}&
\textrm{$S$ (n$\Omega$/mG)}&
\textrm{$\eta_t$ ($\%$)}\\
\hline
TC1N1 & 60 & EP & $27 \pm 13$ & $1.833 \pm 0.004$ & $2.9 \pm 0.6$ & $0.64 \pm 0.06$ & $56 \pm 15$ \\
KEK-R5 & 107 & EP & $26 \pm 10$ & $1.856 \pm 0.004$ & $0.7 \pm 0.1$ & $0.29 \pm 0.01$ & $33 \pm 6$\\
 & & LTB & $27 \pm 13$ & $1.873 \pm 0.004$ & $3.6 \pm 0.3$ & $0.44 \pm 0.02$ & $30 \pm 12$ \\
G2 & 486 & EP &$26 \pm 25$ & $1.867 \pm 0.004$ & $1.8 \pm 0.1$ & $0.59 \pm 0.01$ & $19 \pm 3$\\
 & & N-doping & $26 \pm 25$ & $1.838 \pm 0.004$ & $1.6 \pm 0.2$ & $1.04 \pm 0.01$ & $16 \pm 7$\\
\end{tabular}
\end{ruledtabular}
\end{table*}

\begin{figure}[htb]
\includegraphics*[width=70mm]{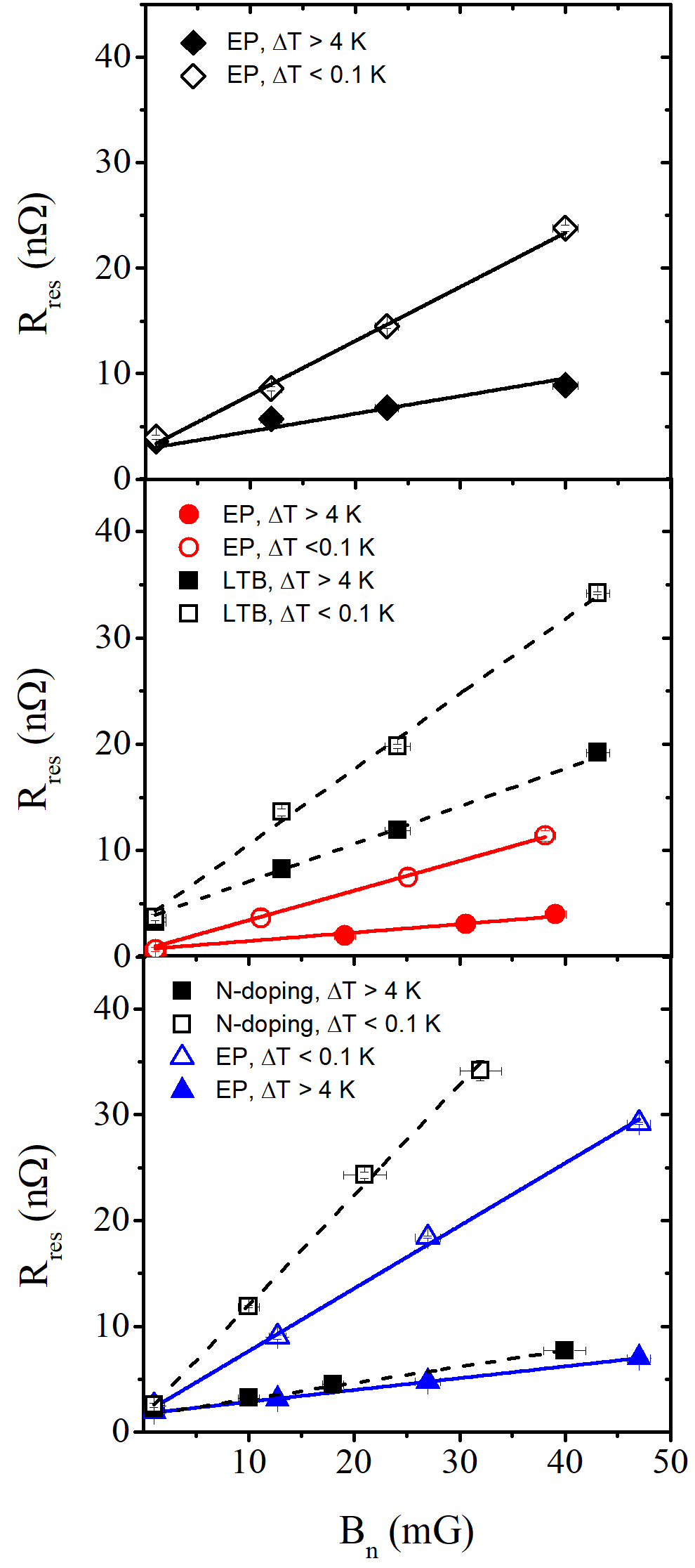}
\caption{\label{fig:RresBa} Residual resistance as a function of applied dc magnetic field, $B_n$, measured for different cool-down conditions and surface treatments for cavities TC1N1 (top), KEK-R5 (middle) and G2 (bottom). The solid lines are linear least-squares fits to the data.}
\end{figure}

Figure~\ref{fig:RresBa} shows the residual resistance as a function of the dc magnetic field before the cavity transitions from the normal to superconducting state, $B_n$, in the two cool-down conditions, one which leads to good flux expulsion ($\Delta T > 4$~K) and one which leads to nearly complete flux trapping ($\Delta T < 0.1$~K). 

For uniform cool-down, the measurements of $B_{sc}/B_n$ indicate that nearly all the magnetic flux is trapped, $B_n \sim B_0$, therefore $R_{res}(B_n)$ can be described by:
\begin{equation}
R_{res}(B_n) = R_{res0} + S B_n,
\label{eq2}
\end{equation}
where $R_{res0}$ accounts for contributions to the residual resistance other than trapped flux, such as nonsuperconducting nano-precipitates, suboxide layer at the surface, broadening of the density of states \cite{AGsust},  etc. For cool-down with large $\Delta T$, only a fraction $\eta_t$ of the applied field is trapped so $R_{res}(B_n)$ can be described by:
\begin{equation}
R_{res}(B_n) = R_{res0} + \eta_t S B_n.
\label{eq3}
\end{equation}
The slope from a least-square linear fit of $R_{res}(B_n)$ for $\Delta T < 0.1$~K is the trapped flux sensitivity, whereas the fraction of the applied field which is trapped can be obtained from the slope of a least-square linear fit of $R_{res}(B_n)$ for $\Delta T > 4$~K. The values of $S$, $R_{res0}$ and $\eta_t$ are listed in Table~\ref{table2} for the three cavities. A common value of $R_{res0}$ was obtained by the least-square fit from the two data sets for each cool-down condition.

The slope of $R_{res}(B_n)$ for cavity G2 after N-doping is close to the value after EP if the cavity is cooled in a large temperature gradient, however it increases by a factor of $\sim 2$ after a uniform cool-down. The residual resistance of cavity KEK-R5 after LTB increased by $\sim3$~n$\Omega$ and $S$ increased by $\sim40\%$, compared to the values after EP. After this set of measurements, the cavity KEK-R5 was re-processed by annealing at 800~\degree C/3 h in a vacuum furnace, followed by $\sim20$ $\mu$m removal by EP and LTB at 120~\degree C/24 h. The measurements of $R_{res}(B_n)$ were repeated and the results were within one standard deviation from the results of the previous test after LTB, indicating the reproducibility of the results.

\begin{figure}[htb]
\includegraphics{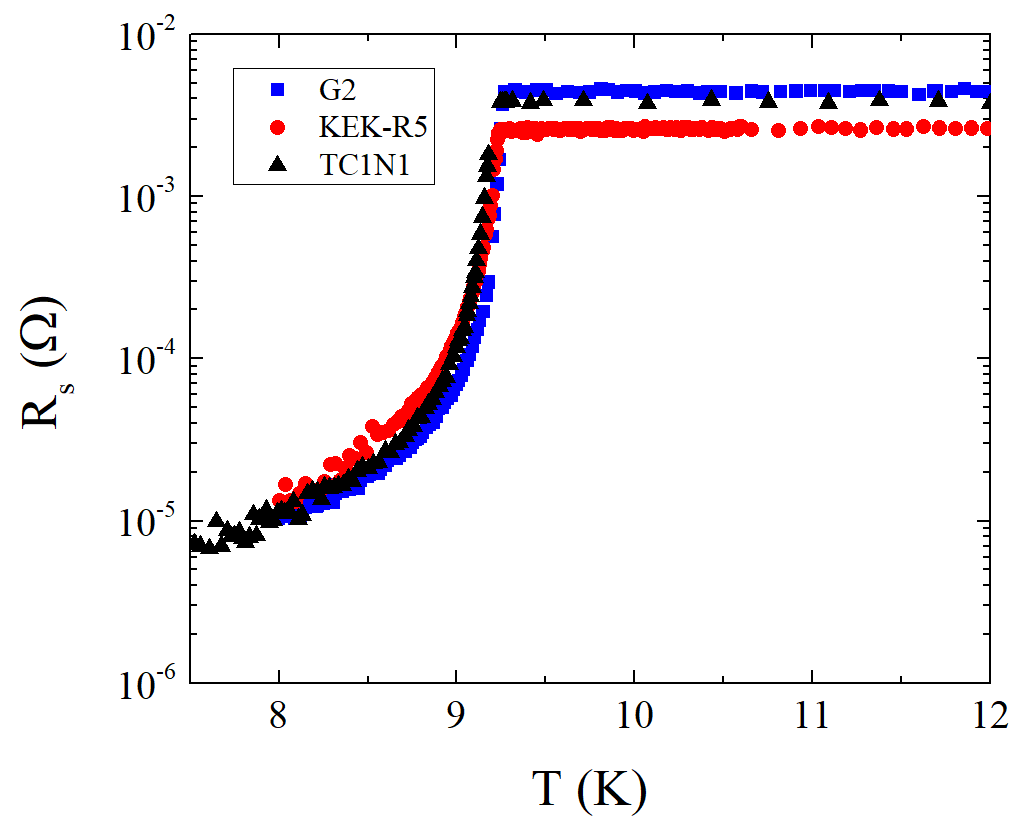}
\caption{\label{fig:Rs} Surface resistance versus temperature between 7.5 K and 12 K measured on cavity TC1N1 after EP, KEK-R5 after LTB and G2 after N-doping.}
\end{figure}

\begin{figure}[htb]
\includegraphics{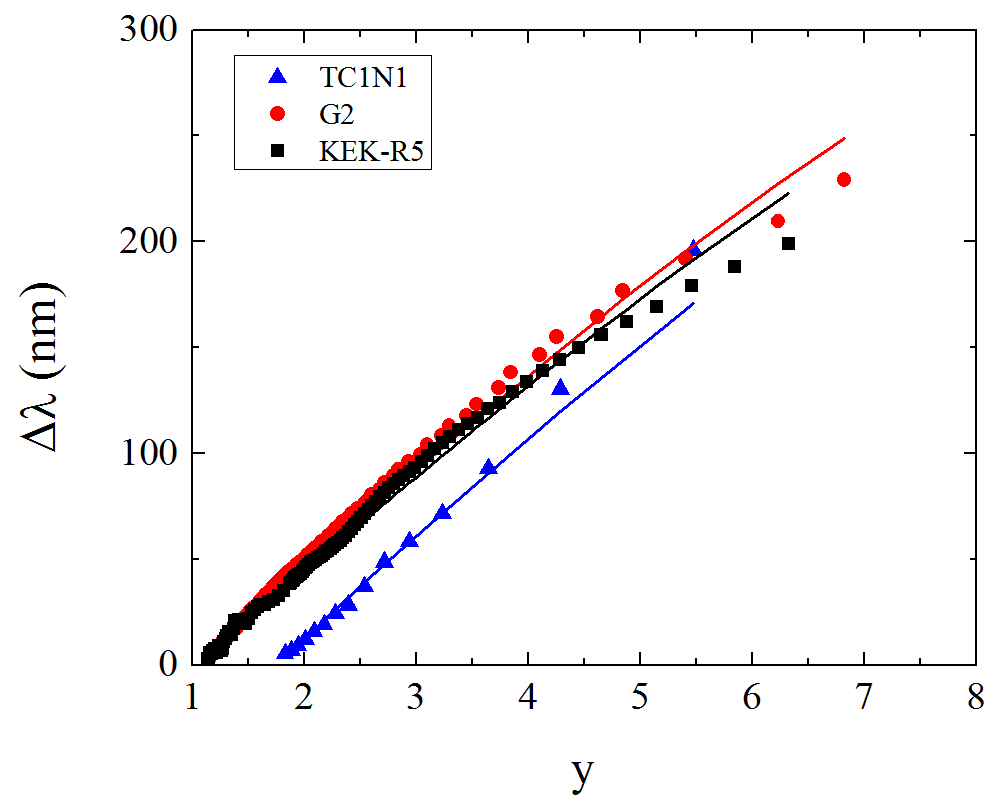}
\caption{\label{fig:lambda} Change of penetration depth as a function of the reduced temperature parameter $y=1/\sqrt{1-(T/T_c)^4}$ measured on cavity TC1N1 after EP, KEK-R5 after LTB and G2 after N-doping. Solid lines are fit with M-B theory.}
\end{figure}

In order to obtain information about the normal state mean free path near the surface, we measured the resonant frequency and the quality factor while warming up the cavities from $\sim5$~K to $\sim10$~K using a vector-network analyzer, from which $R_s(T)$ and the change in rf penetration depth $\Delta \lambda (T)$ can be obtained in this temperature region~\cite{GC_bake}. These measurements were done on cavities TC1N1 after EP, KEK-R5 after LTB and G2 after N-doping at a peak surface rf magnetic field in the range $0.03 - 0.3$~mT and the data are shown in Figs.~\ref{fig:Rs} and \ref{fig:lambda}. The data in the superconducting state were fitted using the numerical solution of M-B theory. The ratio $\Delta /k_BT_c$ was obtained from the fit of $R_s(T)$, whereas $l (7.5 - 9.1$~K) and $T_c$ are weighted averages of the results from the fit of both $R_s(T)$ and $\Delta \lambda (T)$. The normal-state dc resistivity at 10~K, $\rho_n$, was calculated from the value of the surface resistance at 10~K using a numerical solution of the surface impedance of normal metals~\cite{Rn}. To calculate the surface $RRR = \rho(293$~K$)/\rho_n$, we took $\rho(293$~K$) = 14.7~\mu\Omega$ cm. The value of mean free path can be calculated as follows~\cite{AGsust}:
\begin{equation}
\l (10\, \mbox{K})=\frac{\hbar\left(3\pi^2\right)^{1/3}}{n_0^{2/3} e^2 \rho_n},
\label{eq4}
\end{equation}
where $\hbar$ is Planck constant, $e$ is the electron charge and $n_0$ is the electron density. We used $n_0=7 \cdot 10^{28}$~m$^{-3}$ obtained from the measurements of the Hall coefficient $R_H=1/en_0$ in Nb~\cite{Nbn0}.
Table~\ref{table3} lists the values of $T_c$, $\Delta /k_BT_c$ and $l$ from fitting of the surface impedance in the superconducting state, as well as the surface $RRR$, the skin depth, $\delta_n$, and $l$ in the normal state at 10~K.


\begin{table*}
\caption{\label{table3}
Material parameters $T_c$, $l$ and $\Delta /k_BT_c$ obtained from fits of $R_s(T)$ and $\Delta \lambda (T)$ between $7.5-9.2$~K with M-B theory. The surface $RRR$, skin depth, $\delta_n$, and mean free path in the normal state, $\l$(10~K), were obtained from the surface resistance at 10 K.}
\begin{ruledtabular}
\begin{tabular}{ccccccccc}
\textrm{Cavity Name}&
\textrm{Bulk $RRR$}&
\textrm{Treatment}&
\textrm{$T_c$ (K)}&
\textrm{$l (7.5 - 9.1)$~K (nm)}&
\textrm{$\Delta /k_BT_c$}&
\textrm{Surface $RRR$}&
\textrm{$\delta_n$ (nm)}&
\textrm{$l$(10~K) (nm)}\\
\hline
TC1N1 & 60 & EP & $9.19 \pm 0.06 $ & $107 \pm 58$ & $1.90 \pm 0.09$ & 49 & 776 & $254 \pm 7$ \\
KEK-R5 & 107 & LTB & $9.19 \pm 0.05 $ & $122 \pm 74$ & $2.0 \pm 0.2$ & 107 & 573 & $550 \pm 16$ \\
G2 & 486 & N-doping & $9.24 \pm 0.03 $ & $114 \pm 29$ & $1.96 \pm 0.06$ & 37 & 884 & $189 \pm 7$ \\
\end{tabular}
\end{ruledtabular}
\end{table*}

\section{\label{sec:comparison}comparison of experimental data with theoretical models}

Rf dissipation due to trapped vortices has been calculated both for a pinned vortex which has a segment normal to the inner cavity surface~\cite{prb2,AGsust} and for a pinned vortex which has multiple segments parallel to the inner surface~\cite{prb1}, as illustrated by Fig.~\ref{fig:pin_sketch}. Such models allow calculating the trapped flux sensitivity and its dependence on the mean free path and pinning forces. This requires solving the equation of motion of an elastic vortex under the action of the viscous, bending, pinning and the rf current driving forces causing the local displacement of the vortex line $\textbf{u}(z,t)$ in the $xy$ plane \cite{blatter,ehb}:  
\begin{equation}
\eta \dot{\textbf{u}} =\epsilon \textbf{u}''  - \sum_{m}\textbf{f}_p(\textbf{u}-\textbf{r}_m, z-z_m) + \textbf{F} e^{-z/\lambda +i\omega t}.
\label{eq0}
\end{equation}
Here $F=\phi _{0}B_p/\mu_0\lambda$ is the amplitude of rf driving force with the angular frequency $\omega$, $\eta$ is the viscous drag coefficient, the overdot and the prime denote differentiation over time and the coordinate $z$ perpendicular to the surface, respectively.  Eq. (\ref{eq0}) includes the sum of elementary pinning forces $\textbf{f}_p(\textbf{r}-\textbf{r}_m)$ produced by materials defects located at $(x_m,y_m,z_m)$ and the term $\epsilon \textbf{u}''$ accounts for elastic bending distortions. The vortex line tension $\epsilon$ generally depends on the wave number $k_\omega$ of the vortex ripple but for long wavelengths $\lambda k_\omega\ll 1$ relevant to the cases considered below, $\epsilon$ can be approximated by \cite{blatter,lab}
\begin{equation}
\epsilon\simeq \frac{\phi_0^2}{4\pi\mu_0\lambda^2}\left(\ln\kappa+0.5\right),
\label{eps}
\end{equation}  
where $\kappa=\lambda/\xi$ is the Ginzburg-Landau (GL) parameter.

Solution of the nonlinear Eq. (\ref{eq0}) can be simplified using the fact that pinning in Nb cavities is weak, that is, typical depinning critical current densities $J_c$ are orders of magnitude lower than $J_c$ of superconducting materials with artificial pinning centers used in magnets \cite{ce}. For instance, $J_c\sim~10^8$~A/m$^2$ measured on Nb ingots \cite{Dhakal2} is 4 orders of magnitude smaller than the screening depairing current density $J_d\simeq \phi_0/4\pi\mu_0\lambda^2\xi\simeq 2\cdot 10^{12}$ A/m$^2$ which flows at the surface at $H\simeq H_c$. This suggests that pinning may be  produced by either dense arrays of weak materials defects or by sparse arrays of strong pins spaced by distances $\gg \lambda$. In this case calculation of the vortex rf losses can be reduced to the analysis of three representative cases depicted in Fig. 1: 1. A vortex nearly parallel to the surface and pinned strongly by sparse materials defects; 2. A  vortex perpendicular to the surface and pinned strongly by a materials defect spaced by $\ell$ from the surface. 3: A vortex perpendicular to the surface pinned collectively by randomly distributed weak defects. Calculations of $R_{res}$ for these cases were done in Refs. \cite{prb1,prb2,AGsust}. The corresponding formulas used for the analysis of the experimental data are given in the Appendices. Here we do not consider strong correlated pinning caused by planar grain boundaries like in Nb$_3$Sn of $\alpha$-Ti ribbons like in NbTi, or columnar defects like dislocations or radiation tracks \cite{blatter,ehb,GB,agcond,anl}.  

For vortices parallel to the surface  \cite{prb1}, the main contribution to $R_{res}$  comes from vortex segments exposed to the rf current. The distance of the vortex from the surface $d$ cannot be shorter that a critical value $d_m$ at which the attraction of the vortex to the surface exceeds the pinning force.  As shown in Appendix~\ref{App:B},  $d_m$ is determined by the following equation:
\begin{equation}
e^{-2d_m/\lambda}\simeq \frac{\kappa J_c}{J_d}\sqrt{\frac{d_m}{\pi \lambda}},
\label{dm}
\end{equation}
where $J_c\sim f_p/\ell\phi_0$ is a depinning current density, $\ell$ is a mean pin spacing, and $f_p$ is an elementary pinning force:
   \begin{equation}
   f_p = \zeta \mu_0 \pi H_c^2 \xi^2.
    \label{f_pl}
    \end{equation}
The parameter $\zeta$ which quantifies the gain in the condensation energy at the pinning defect is maximum $(\zeta\sim 1)$  for the strongest core pining by a dielectric precipitate of radius $r_0\simeq \xi$ \cite{ce,blatter,ehb,anl}. For a small precipitate $r_0<\xi$, we have $\zeta \sim (r_0/\xi)^3\ll 1$ \cite{ce,blatter}. For atomic impurities, $\zeta\sim \sigma/\xi^2\ll 1$, is proportional to the scattering cross-section $\sigma$ on the impurity in the normal state \cite{thun}. 

If the vortex stretched along the applied dc field $\textbf{B}_0$ \cite{b} 
gets within the expulsion distance $d_m$ from a curved cavity surface, it splits into 
two disconnected parts, as shown in Fig. 1. The parallel vortex segments exist within a belt of width $h$ along the equator, where $h$ can be evaluated from the condition $R^2\simeq (R-d_m)^2+h^2$. Hence, 
$h\simeq (2Rd_m)^{1/2}$, where $R$ is the curvature radius of the cavity and $d_m\sim (\lambda/2)\ln(J_d/\kappa J_c)$ follows from Eq. (\ref{dm}) with a logarithmic accuracy in $\ln^{-1}(J_d/J_c)\ll 1$. Thus, 
\begin{equation}
h\sim\sqrt{\lambda R}\ln^{1/2}\frac{J_d}{\kappa J_c}.
\label{hh}
\end{equation}
For $R\simeq 0.1$ m, $\lambda=40$ nm, $J_c\sim 10^{-3}J_d$, and $\kappa=1$, Eq. (\ref{hh}) gives $h\sim 0.1$ mm.
The flux sensitivity is then $S_\|\sim R_{res}\gamma_\|/B_0$, where $\gamma_\|= 2\pi h R/A$ is the fraction of the cavity surface area $A$ contributing to the trapped flux losses. Our calculations of $S_\|$ using the formulae for $R_{res}(B_0)$~ \cite{prb1} at $\gamma_\|\sim 10^{-4}$ have shown that pinned vortex segments parallel to the surface result in $S_\|$ which is some 4 orders of magnitude smaller than $S$ values observed on Nb cavities. Therefore, the main contribution to the flux sensitivity in Nb cavities comes from trapped vortices perpendicular to the surface \cite{prb2}. Two essential contributions to $S$ are considered below.

For a perpendicular vortex pinned by a strong defect at $z=\ell$, the trapped flux sensitivity is given by \cite{prb2}:
\begin{gather}
S\simeq \frac{\gamma\phi_0\chi^2}{2\eta\lambda}\biggl[\frac{5+\chi^2}{(1+\chi^2)^2}-\frac{2}{\chi^{3/2}}\mbox{Im}\frac{\tanh\sqrt{i\nu}}{\sqrt{i}(1-i\chi)^2}\biggr],
\label{S_norm}\\
\chi=\frac{\omega\eta}{\varepsilon}\lambda^{2}, \qquad \nu=\frac{\omega\eta}{\varepsilon}\ell^{2}.
\label{nu}
\end{gather}
Here the rf current causes bending disturbance extending over the ripple length $L_\omega$ along the vortex line \cite{prb2}
\begin{equation}
L_\omega=\sqrt{\frac{\epsilon}{\eta\omega}}=\frac{\xi}{2\lambda}\sqrt{\frac{g\rho_n}{\pi\mu_0f}},
\label{Lo}
\end{equation}
where $g=\ln(\lambda/\xi)+1/2$, $f$ is the rf frequency and $\eta=\phi_0^2/2\pi\xi^2\rho_n$. For Nb with $\lambda\approx\xi$ and $\rho_n=10^{-9}$~$\Omega$~m, we have $L_\omega\simeq 180$ nm at 1 GHz. Thus, oscillating bending distortions of the elastic vortex can extend well beyond the rf field penetration depth, $L_\omega$ being practically independent of $T$ and decreasing as the m.f.p. decreases. For instance, in the dirty limit, $\lambda\simeq\lambda_0(\xi_0/l)^{1/2}$ and $\xi\simeq \sqrt{\xi_0 l}$, we have $L_\omega ^{dirty}\simeq  L_\omega^{clean}\sqrt{l/\xi_0}$. If the pin distance $\ell$ exceeds $L_\omega$ the flux sensitivity is independent of the pinning force.  The factor $\gamma$ in Eq. (\ref{S_norm}) takes into account the spatial distribution of trapped vortices over the cavity surface (see Appendix B):
\begin{equation}
\gamma=\frac{\phi_0\int n(\textbf{r})H^2(\textbf{r})dA}{B_n\int H^2(\textbf{r})dA},
\label{G}
\end{equation}
where $n(\textbf{r})$ is the local areal density of vortices coming out of the inner cavity surface, $H(\textbf{r})$ is a tangential component of RF magnetic field at the surface. For a statistically-homogeneous distribution of trapped vortices piercing the cavity along the dc field $\textbf{B}_n$, the ratio $n(\textbf{r})\phi_0/B_n\to\cos\theta(\textbf{r})$ depends only on the angle $\theta(\textbf{r})$ between the normal to the cavity surface and $\textbf{B}_n$, and $\gamma$ depends only on the cavity shape and the rf mode.  For the TM$_{010}$ mode in the elliptical cavities studied in this work, we calculated $\gamma=0.61$ numerically and used this value in the analysis of the experimental data.  

Since the amplitude of the rf ripples along the vortex line decreases exponentially over the length $L_\omega$, pins spaced by $\ell\gtrsim L_\omega$ from the surface have no effect on $R_{res}$, whereas pins closer to the surface reduce $R_{res}$. For sparse pins,  $R_{res}$ is dominated by dissipative oscillations of free vortex segments between the pins. Thus, vortex segments of length $\ell>L_\omega$ cause the highest rf losses independent of details of the pinning forces $f_p(u)$. The net rf power is determined by statistical averaging of $R_{res}$ over the random pin spacing from the surface \cite{prb1,prb2}
\begin{equation}
\bar{R}_{res}=\int_0^\infty G(\ell)R_{res}(\ell)d\ell,
\label{br}
\end{equation}
where $G(\ell)$ is a distribution function of the pin spacings.  Random distribution of the nearest pin positions along the cavity surface manifests itself in strong fluctuations of local vortex losses in hotspots caused by vortex bundles pinned deep inside the cavity wall and having long dangling segments of length $\gtrsim L_\omega $ at the surface.

In the case of weak collective pinning of a perpendicular vortex, Eq. (\ref{eq0}) can be simplified to the following equation for small displacement of the vortex $u(z,t)$:
\begin{equation}
\eta \dot{u}=\epsilon u'' -\alpha u + F e^{-z/\lambda +i\omega t}.
\label{eq5}
\end{equation}
Here the term $-\alpha u$ describes the effect of pinning, and the Labusch spring constant $\alpha$  \cite{ce,blatter,ehb,lab} is evaluated in Appendix C for arrays of small nanoprecipitates or atomic impurities. The rf current flowing at the surface causes oscillations of the vortex line which extend over the complex Campbell penetration length \cite{ce,ehb,lab}:  
\begin{equation}
\lambda_c=\left[\frac{\epsilon}{\alpha+i\omega\eta}\right]^{1/2}.
\label{camp}
\end{equation}
For weak collective pinning at GHz frequencies, $\omega\eta\gg \alpha$, Eq. (\ref{camp}) reduces to Eq. (\ref{Lo}), giving $L_\omega\to\lambda_c$ which can significantly exceed $\lambda$, as it was shown above. In this case the rf losses occur not only in the surface layer of the rf currents but also come from oscillations of long segments of vortex lines extending deep inside the cavity wall over the length $\sim L_{\omega}\gg \lambda$.  In the static limit $\lambda_c$ reduces to the Larkin pinning correlation length $L_c\simeq \xi(J_d/J_c)^{1/2}$~ which defines a length scale of bending distortion along the vortex line  \cite{blatter,ehb,ag_pin}. For Nb with $\lambda\approx\xi\approx 40$ nm, $J_d\simeq \phi_0/4\pi\mu_0\lambda^2\xi= 2\cdot~10^{12}$~Am$^{-2}$ and  $J_c\sim 10^8-10^9$~Am$^{-2}$ ~\cite{Dhakal2,ce}, we have $J_c\sim (10^{-4}-10^{-3})J_d$ and $L_c\simeq 2-4~ \mu$m, consistent with the fits of experimental data presented below. 

The trapped flux sensitivity in the case of weak collective pinning is given by (see Appendix B):
\begin{equation}
S(\omega, l)=-\frac{\gamma\phi_0 \chi}{2 \eta \lambda}\mbox{Im} \left[\frac{s+2}{s(s+1)^2}\right],
\label{S_wcp}
\end{equation}
where $s=\lambda/\lambda_c=\sqrt{k+i\chi}$ and $k=\alpha\lambda^2/\epsilon$. 

In this work we focus on trapped vortex losses at low fields, $H_p\ll H_c\simeq 200$ mT, leaving aside complex issues of nonlinear vortex losses at high fields.  Low-frequency vortex losses at high rf fields have been addressed theoretically both for weak collective pinning and hysteretic depinning of vortices from strong pins \cite{lab}.  A quasi-static theory of collective pinning was used to address the linear dependence of the vortex surface resistance on the rf field amplitude \cite{corn} observed on Nb cavities. In what follows we use Eqs. (\ref{S_norm}) and (\ref{S_wcp}) to reveal manifestations of different pinning mechanisms in the observed dependencies of $S$ on the mean free path in Nb cavities. 

\subsection{\label{sec:mfp-dependence}Mean free path dependence}

The electron mean free path can be altered by surface treatments so getting the values of $l$ from the cavity measurements is not always straightforward. Usually $l$ is extracted from fitting the observed surface impedance, $Z_s(T)$, using numerical solutions of the M-B theory \cite{MB,Halb}. The $l$ values can vary depending on the temperature-dependent depth probed by rf current \cite{GC_bake}. There are many uncertainties in evaluating $l$ from the M-B fits coming from both the BCS model assumptions and/or computational intensive grid-search methods to find a global minimum of chi-squared~\cite{cern}. Additional contributions to the rf losses can result from a proximity coupled thin suboxide metallic layer \cite{gk,kg}, common broadening of the gap peaks in the idealized BCS density of state \cite{AGsust}, significant effects of strong electron-phonon coupling in Nb \cite{carbotte} or two-level systems ~\cite{Roman3} which are not taken into account in the M-B model. By contrast, obtaining $l$ from the Drude Eq. (\ref{eq4}) only requires knowledge of $\rho_n$ and the electron density. 

At GHz frequencies the normal skin depth  is about $3-10$ larger than the rf penetration depth at $T < 0.85T_c$, so measurements of $Z_n(T,f)$ in the normal state probe a thicker surface layer across which $l$ can vary due to materials treatment. However, measurements of $Z_n(T)$ at microwave frequencies on Nb coupons at $T$ slightly above $T_c$ may give a more reliable information about $l$ in the $40-100$ nm thick surface layer relevant to SRF cavities. Evidences of variation of $l$ across the surface were   obtained from muon spin rotation ($\mu$SR) experiments on Nb samples treated by EP and LTB which showed that $l$ changes from $l\simeq 2$ to $16$~nm within the depth in the 100 nm surface layer~\cite{Roman2}.  The M-B fit of $R_s(T)$ in LTB cavities gave $l\sim26$~nm and $l\gtrsim 200$~nm in cavities treated by EP~\cite{GC_bake}. The values of $l(1.5-4.3$~K) $\sim26$~nm extracted fom the M-B fits for all three EP cavities shown in Table~\ref{table2} are lower than typical, which may be due to the cavities' treatment history. 
 
 \begin{figure}[htb]
\includegraphics{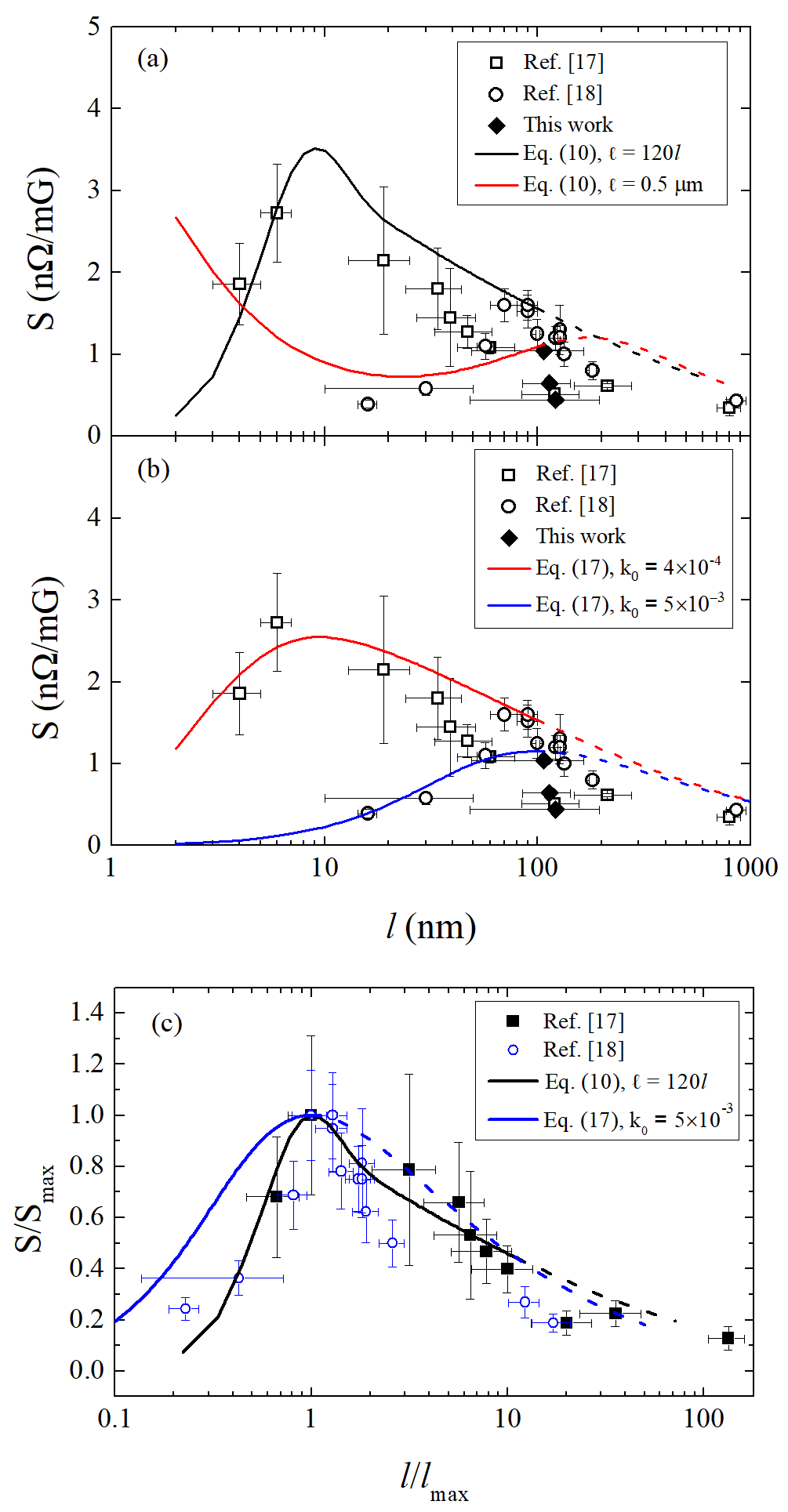}
\caption{\label{fig:S_mfp} Trapped flux sensitivity at 1.3 GHz as a function of mean free path. Solid lines are calculated for the case of a vortex normal to the surface pinned by a strong pin (a) or by weak collective pinning (b). Dashed lines are extrapolations to the clean limit for each case. The mean free path values were obtained from M-B fits above 5 K using Halbritter's code. The trapped flux sensitivity data from Refs.~\cite{DG1} and \cite{FnalS} are shown normalized to the respective peak value and plotted as a function of the mean free path scaled to the value at which $S$ is maximum in each data set in (c). }
\end{figure}
  
Our flux sensitivity data plotted as a function of $l$ are shown in Fig. \ref{fig:S_mfp} (a) and (b). To see how different cavity treatments manifest themselves in the observed flux sensitivity, we also plotted the $S(l)$ data observed on fine-grain 1.3 GHz cavities made of high-purity Nb and of the same shape as the cavities in our work  \cite{FnalS, DG1}. In Ref. \onlinecite{FnalS} fifteen different cavities were subjected to different annealing treatments followed by EP.  In Ref. \onlinecite{DG1} six different cavities were subjected to different annealing followed by different amount of material removal by EP. The $S$-values of Ref. \onlinecite{DG1} were multiplied by a correction factor of 0.58 ~\cite{ML2} 
to be compared with the data of Refs.~\onlinecite{FnalS, FnalF} and our work. As far as we are aware, all flux sensitivity data shown in Fig. \ref{fig:S_mfp}  have been obtained using the same experimental methodology, where the $l$ values have been extracted from the M-B fits of the temperature-dependent surface impedance \cite{MB,Halb}. 

Both $S(l)$ data sets of Refs. \onlinecite{FnalS, DG1} have clear maxima as functions of the m.f.p. but with very different  
values of the peak position and magnitude, $l_{max}$ and $S_{max}$. This indicates that different treatments of cavities done in  Refs. \onlinecite{FnalS, DG1} 
produce different distribution and type of pinning centers which manifest themselves in different flux trapping efficiency and rf  losses. However, if the flux sensitivity data of Refs. \onlinecite{FnalS, DG1} are normalized to their respective values of $S_{max}$ and plotted as  functions $l/l_{max}$, both datasets approximately collapse 
onto a universal bell-shape curve, as shown in Fig. 11 c. This behavior suggests a scaling of $S(l)$ which will be discussed later. Strong scatter of experimental data likely results from the fact that all data points in Fig.  \ref{fig:S_mfp} correspond to different cavities which have undergone different  
treatments resulting in particular values of $l$. However, such materials treatment can not only change the mean free path but also spatial distribution, strength and volume density of pins, so the data in Fig. 11 represent a convoluted effect of materials treatments on both $l$ and pinning characteristics.  Indeed, our experimental data exhibit significant changes in the S-values but rather small changes in $l$ after the cavity treatments, indicating that they mostly affected pinning characteristics rather than the mean free path. 

Now we relate the correlation of $S$ and $l$ observed Refs. \onlinecite{FnalS, DG1} and shown in Fig. \ref{fig:S_mfp} to different pinning mechanisms.  Consider first  the rf losses caused by perpendicular vortex segments pinned by strong sparse pins.  Shown in Fig.~\ref{fig:S_mfp}(a) is $S(l)$ calculated from  Eq. (\ref{S_norm}) at $1.3$ GHz using $\ell$ as a fit parameter and the dependencies of $\lambda(l)$, $\xi(l)$, $\eta(l)$ on mean free path from Appendix~\ref{App:A}. We used the vortex viscosity $\eta(l)=\phi_0^2/2\pi\xi^2\rho_n$  given by the Bardeen-Stephen model, although this model is valid only in the dirty limit $l\lesssim \xi_0$. Therefore, a discrepancy between the theory and experimental data can be expected in a moderate clean limit $l\gtrsim \xi_0$, where the results of calculations are shown as dashed lines in Fig.~\ref{fig:S_mfp}. Here $S(l)$ calculated at a fixed $\ell$ has a broad maximum at $l\simeq 100$ nm.  The maximum in $S(\ell)$ becomes more pronounced if the pin spacing $\ell$ is proportional to $l$, as it was proposed in \cite{DG1}. This assumption might be justified if pinning is caused by small precipitates which also act as electron scattering centers.  

Figure~\ref{fig:S_mfp}(b) shows the fits of the flux sensitivity data to the model in which $S(l)$ is caused by perpendicular vortices pinned collectively by weak small pins. Here $S$ was calculated from Eq. (\ref{S_wcp}) at $1.3$ GHz and $\gamma=0.61$, using the Labusch spring constant $\alpha(l)=\alpha_0(1+\xi_0/l)$ evaluated in Appendix C and  regarding the pinning parameter $k_0=\alpha_0\lambda_0^2/\epsilon_0$ in the clean limit as a fit parameter.
 As $l$ decreases, a maximum in $S(l)$ occurs due to interplay in the decrease of the vortex viscosity $\eta(l)$ in a moderately clean limit and the increase of the pinning constant $k(l)=\alpha\lambda^2/\epsilon\simeq k_0(1+\xi_0/l)^3$ as the vortex line gets softer in the dirty limit. 
 
As follows from Fig.  \ref{fig:S_mfp}, perpendicular trapped vortices pinned by either strong sparse pins or by collective interaction with random array of weak pins can result in bell-shape $S(l)$ dependencies, in qualitative agreement with experiments.  The fits are hardly perfect, which may reflect the fact that both $l$ and pinning characteristics are generally affected by the cavity treatments. In addition, several pinning mechanisms operating on different scales can contribute to $S$, the relative weight of these contributions can vary along the cavity surface.  For instance, vortex hotspots can occur either in regions devoid of strong pins or regions with weak $\delta l$ pinning due to impurity fluctuations or regions with weak random $\delta T_c$ pinning. Networks of dislocations can be clustered in some regions of the surface and be absent in others.  Because of very low densities of vortices in cavities, lateral fluctuations of pinning along the cavity surface are very strong, and different pinning mechanisms can operate simultaneously in different hotspots. Here the largest contributions to $S$ likely come from regions with weak or no pinning in the first $100-200$ nm at the inner surface of the cavity. 

If $S$ is mostly determined by the weak collective pinning, the value of the parameter $k_0=\alpha_0\lambda_0^2/\epsilon_0$ used to fit the $S(l)$ data in Fig.~ \ref{fig:S_mfp} (b) allows us to roughly evaluate pinning characteristics.  For small dielectric nanoprecipitates of radius $r_0<\xi_0$, the mean pin spacing $\ell$ can be expressed in terms of $k_0$ using Eq. (\ref{k0}):
\begin{equation}
\ell\simeq\frac{\lambda_0}{\sqrt{k_0}}\left(\frac{2}{3g}\right)^{2/3}\!\left(\frac{r_0}{\xi_0}\right)^{2},
\label{l1}
\end{equation}
where $n_p=\ell^{-3}$ is the volume density of pins.  
For instance, if $r_0=5$ nm, $\xi_0=\lambda_0=40$ nm, and $g=\ln\kappa+1/2=1/2$, Eq. (\ref{l1}) yields 
$\ell\simeq \lambda_0\simeq 38$ nm at $k_0=4\cdot 10^{-4}$. Weaker proximity coupled metallic nanoprecipitates require a higher pin density as compared to dielectric precipitates to provide the same value of $k_0$. On the other hand, Eq. (\ref{l1}) may   
overestimate the volume density of nanoprecipitates because 
another contribution to $\alpha$ comes from $\delta l$ pinning due to fluctuations of the mean free path. As shown in Appendix C, 
$\delta l$ pinning can be essential in a dirty surface layer with sparse small nanoprecipitates if the condition (\ref{ineq}) is satisfied. 

The Labusch constant $\alpha$ can be expressed in terms of a depinning current density $J_c$ by equating the Larkin pinning correlation length $L_c\sim \xi\sqrt{J_d/J_c}$ \cite{blatter} to the static Campbell length $\lambda_c=\sqrt{\epsilon/\alpha}$, where $J_d=\phi_0/4\pi\mu_0\lambda^2\xi$ and $\epsilon$  is given by Eq. (\ref{eps}). Hence,
\begin{equation}
\alpha\simeq \phi_0gJ_c/\xi.
\label{alph}
\end{equation}
It is instructive to express $J_c$ and $L_c$ in terms of the dimensionless pinning parameter $k=\alpha\lambda^2/\epsilon$ extracted from the fits of Eq. (\ref{S_wcp}) to the flux sensitivity data:
\begin{equation}
J_c=kJ_d/\kappa^2,\qquad L_c=\lambda/\sqrt{k}.
\label{cpt}
\end{equation}  
For the typical values of $k_0\sim 10^{-3}$, we obtain 
$J_c\sim 10^{-3}J_d$ and $L_c\sim 30\lambda$ for clean Nb with $\kappa\simeq 1$. This shows that: 1. Pinning in Nb cavities is indeed weak, as was mentioned above. 2.  Dissipative oscillations of the elastic vortex extend well beyond the layer of the surface rf current which excites these oscillations. Notice that these $J_c$ values correspond to a layer $z\lesssim \lambda_c$ at the Nb surface where the density of structural defects which can pin vortices is typically much higher than in the bulk. Because of stronger pinning in the surface layer caused by different materials and mechanical treatments used in the cavity production, it is not surprising that the $J_c$ values extracted from the $S$-fits are an order of magnitude higher than global $J_c$ obtained from the measurements of magnetization loops on Nb ingots \cite{Dhakal2}.

\subsection{\label{sec:f-dependence}Frequency dependence}
\begin{figure}[htb]
\includegraphics{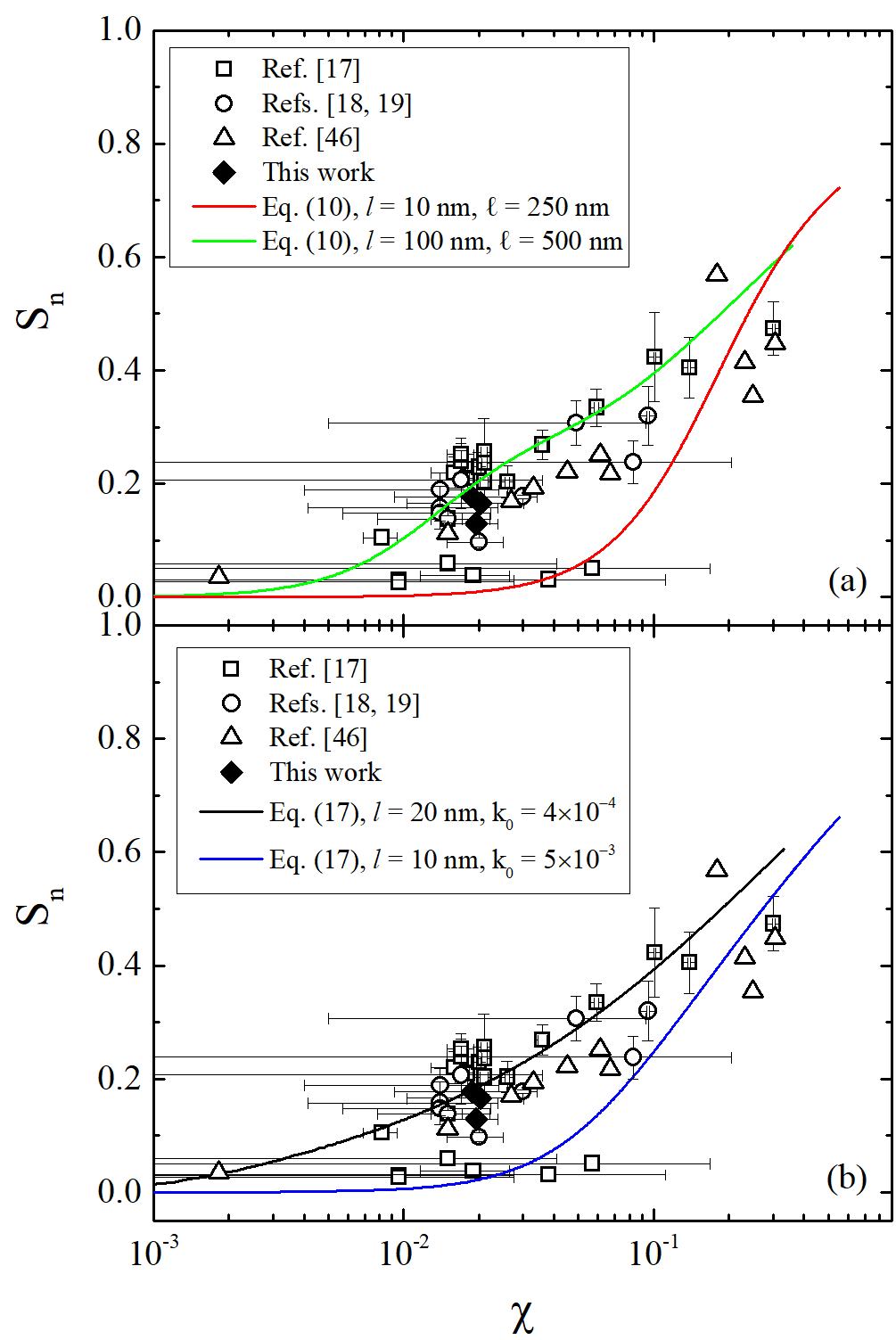}
\caption{\label{fig:S_f} Normalized trapped flux sensitivity as a function of the dimensionless frequency $\chi = \omega \eta \lambda^2/\epsilon$ for a vortex normal to the surface by a strong pin (a) or weak collective pinning (b). Solid lines are calculated with Eqs.~(\ref{S_norm}) and (\ref{S_wcp}) with different values of models parameters shown in the legend.}
\label{S_f}
\end{figure}

Figure~\ref{fig:S_f} shows the trapped flux sensitivity normalized to the respective high-frequency limits $S_n = S/S_{hf}$ as a function of the dimensionless frequency $\chi=\omega\eta\lambda^2/\epsilon$. The data from Refs. \onlinecite{FnalS, DG1, FnalF, Vallet} are plotted along with the data from this work and are fitted to Eqs.~(\ref{S_norm}) and (\ref{S_wcp}) for different values of the model parameters.

Since $\chi$ depends on $l$, cavities resonating at the same frequency but with different mean free path values result in different values of $\chi$. While most of the data shown in Fig.~\ref{fig:S_f} are for 1.3~GHz cavities, there are also three data points each for 650~MHz, 2.6~GHz and 3.9~GHz elliptical cavities~\cite{FnalF}. The data in Ref.~\onlinecite{Vallet} were obtained from different cavities with frequencies in the range 81 MHz - 21.5 GHz and had been corrected for the cavity geometry with respect to the direction of the applied field.  

It should be pointed out that in Refs.~\onlinecite{FnalS, FnalF} the mean free path in cavities after LTB was not obtained from RF measurements but the same value $l = 16$~nm was assigned, based on $\mu$-SR measurements on Nb coupons. The mean free path close to the surface was also not measured in Ref. \onlinecite{Vallet} and we used the value obtained from the reported bulk RRR in order to calculate $S_{hf}$ and $\chi$ for such set of data. These assumptions along with the issues discussed above  can contribute to the strong scatter of the data in Fig. ~\ref{fig:S_f}.

\section{\label{sec:disc}Discussion}

\subsection{Bulk vs. surface pinning}
The data listed in Tables~\ref{table2} show that the fraction of magnetic field trapped during cool-down with $\Delta T > 4$~K increases with decreasing bulk $RRR$ of the cavity and it is not significantly affected by surface treatments, such as N-doping and LTB. This important finding suggests that pinning is dominated by the bulk materials properties, which is consistent with the results of Ref.~\onlinecite{PosenTF}. The grain structure is similar in all three cavities, and the major differences are in the concentration of interstitial impurities, which should be uniformly distributed in the material. On the other hand, the trapped-flux sensitivity does not seem to be correlated with the bulk $RRR$. This can be expected since only trapped-vortex segments at the surface contribute to rf losses. The flux sensitivity $S$ increased by $\sim50\%$ after LTB and $\sim76\%$ after N-doping, showing that surface treatments significantly affect $S$, consistent with published data on fine-grain Nb cavities \cite{DG1,FnalS,PosenTF}.
Lower material purity can result in a larger fraction of the trapped flux because the vortex line tension $\epsilon\simeq \epsilon_0(1+\xi_0/l)^{-1}$ decreases as $l$ decreases, therefore making it easier for a vortex to be pinned. While the surface RRR is similar to the bulk value for the low-purity cavity after EP and the medium purity cavity after LTB, it is much smaller than the bulk value for the high-purity cavity after N-doping. This suggests that the diffusion of N during the infusion process occurs over a depth of the order of the skin-depth $\sim 1$~$\mu$m in this case. This result is consistent with measurements of the impurities depth profile in N-doped Nb samples \cite{AnnaSRF15, Tuggle}.

The scatter of the $S(l)$ data shown in Figs. 11 and 12 likely comes from the measurements of $S$ on different cavities which have undergone different treatments affecting both $l$ and pinning characteristics. Another contribution to the scatter of the $S$-data may come from the ambiguity in determining $l$ within the top $\sim 40$~nm surface layer, as it was discussed in Sec. \ref{sec:mfp-dependence}. The data from Refs.~\cite{FnalS, DG1} exhibit a maximum in $S(l)$ but the position of the maximum $l_m$ in these two sets of data differ by an order of magnitude. Such a big difference in the peak positions $l_m$ can hardly be entirely attributed to different ways of extracting $l$ from the data used by different groups but rather indicates a significant difference in pinning strengths which causes flux trapping in the first place. After rescaling $S/S_{max}$ as a function of $l/l_{max}$ both datasets approximately collapse onto a universal curve as shown in Fig.  \ref{fig:S_mfp} (c). Such scaling is indicative of $S(p)$ being a function of one parameter $p$ which absorbs both the mean free path and pinning characteristics.  The single-parameter scaling takes place in $S(s)$ given by Eq. (\ref{S_wcp}) for the collective pinning model and $S(\nu)$ in Eq. (\ref{S_norm}) for strong sparse pins in the limit of $\chi\ll 1$ characteristic of Nb cavities (see Fig. \ref{S_f}).

Analyzing $S(l)$ as a function of only the mean free path does not give a complete picture of flux sensitivity since heat treatments change not only $l$ but also spatial distributions of impurities or oxide/hydride nanoprecipitates affecting the $\delta\ell$ or $\delta T_c$ collective pinning, or correlated pinning by grain boundaries or dislocation networks. For instance, our data which show a significant change in flux sensitivity without much change in $l$ could be understood assuming that LTB or N doping facilitate either a diffusive coalescence or dispersion of pinning nanoprecipitates.  In this case both the radius of precipiates $r_0$ and the pin spacing $\ell$ would change after each heat treatment but the volume fraction of a nonsuperconducting phase $4\pi r_0^3n_p/3=(4\pi/3)(r_0/\ell)^3$ remain constant. As a result, the pinning parameter $k_0$ which controls the behavior of $S(l)$ for the collective $\delta T_c$ pinning would evolve as $k_0\propto r_0^2$ at a constant ratio $r_0/\ell$ in Eq. (\ref{k0}). Cavity treatments could cause diffusive shrinkage or dissolution of sparse strong pinning oxide or hydride nanoprecipitates within the layer of thickness $\simeq L_\omega$  at the surface, which would obviously increase flux losses.  Heat treatments can also affect segregation of impurities on dislocations or grain boundaries which changes pinning forces \cite{GB}. 

\subsection{Models of the trapped flux sensitivity and comparison with experimental data}

The flux sensitivity $S(l,f)$ calculated for different pinning mechanisms increases as $l$ decreases in the clean limit and then decreases at shorter $l$, as shown in Fig.~\ref{fig:S_mfp}.  Here the maximum in $S(l)$ results from interplay of the decrease of the vortex viscosity and increase of the pinning strength as $l$ decreases as was also pointed out by Checchin et al. \cite{Fnal_sust} using the Gittleman-Rosenblum (GR) model \cite{gr}.  We showed that the observed $S$-values can be obtained from Eq.~(\ref{S_norm}) with $\ell \gg \lambda_0$ or from Eq.~ (\ref{S_wcp}) with reasonable assumptions regarding the size and concentration of defects. The position and the magnitude of the maximum of $S(l)$ depend on frequency and the pinning strength quantified by either the Labusch constant $\alpha$ of the pin spacing $\ell$. The main contribution to $S$ comes from weak pinning regions in which dissipative oscillations along vortices extend into the bulk well beyond the layer of rf screening current.

Identifying a correlation between the materials defects and vortex hotspots would require rf measurements of $Q(T)$ and temperature maps on the cavity combined with electron microscopy of  coupons cut out from the same cavity.  Yet even such putative state-of-the-art experiments may not pinpoint the pinning defects responsible for the dominating rf losses of trapped flux. Indeed, the strongest hotspots are produced by perpendicular vortex segments which are either pinned collectively by many weak pins such as clusters of atomic impurities or by strong pins like nonsuperconducting precipitates, which can be hundreds nanometers away from the surface. In that case surface probes can miss the materials defects resulting in the strongest vortex losses. 

Models of pinned vortices driven by the rf current must include a finite vortex line tension $\epsilon$, otherwise there would be no pinning \cite{ce, blatter,ehb,agcond,anl}. Indeed, at $\epsilon\to\infty$ a straight stiff vortex cannot be pinned by randomly distributed  materials defects. In the limit of zero $\epsilon$, soft vortex segments between pinning centers would bow out and reconnect under any infinitesimal Lorentz force of current  \cite{ehb,agcond,anl}. These issues are relevant to the models of rf vortex losses ~\cite{Fnal_sust, calatroni} in which the vortex line tension was disregarded and the GR model \cite{gr} was used. However, the GR model was proposed to describe short perpendicular vortices driven by a uniform rf current in a thin film. By contrast, a rf current flowing at the cavity surface excites dissipative ripples along the elastic vortex extending over the length $L_\omega \gg \lambda$ and producing losses deep inside the cavity wall.    
This effect cannot be described by the GR model which is applicable to nearly straight perpendicular vortices in thin films or SRF thin coatings, if the film thickness is smaller than the Campbell penetration depth $\lambda_c$ \cite{ag_pin}. In the static limit $\lambda_c$ becomes the Larkin pinning correlation length $L_c\sim 10^2\lambda$ estimated above. At 2 GHz $L_\omega$ drops down to $\simeq 200$ nm which is about 5 times larger than $\lambda$ in clean Nb.

The randomness of spatial distribution of pinning centers in the cavity wall can result in strong local fluctuations of flux trapping efficiency which are especially pronounced at  low density of trapped vortices. The flux sensitivity $S$ then results from spatial averaging over distributions of $\ell$-values, pinning strengths and positions of vortex segments relative to the surface. However, $S$ also depends on such extrinsic factors as the history of cooling the cavity through $T_c$, temperature cooling rate and the directions and magnitudes of local temperature gradients. Moreover, local flux losses can vary significantly even if the average densities of pinning centers and trapped vortices are constant. Indeed, the regions which have pinning centers within the rf surface layer would greatly reduce flux losses, whereas pin free regions at the rf surface would have much stronger flux losses if the vortex is trapped by pins deep inside the cavity wall.

\subsection{Optimization of flux pinning}

Given the multitude of mechanisms of flux trapping and their strong dependencies on the materials treatment, one could pose the question: Is flux trapping inevitable and to what extent can the vortex losses be reduced to an acceptable level by optimizing pinning nanostructure? The answer to the first part of the question is yes: flux trapping occurs during the cavity cooldown through $T_c$ at which the energy barrier for the vortex creation vanishes so any materials defects both in the rf layer and deep inside the cavity wall can trap vortices. A fraction of these vortices escapes upon cavity cooldown but some of them remain trapped. The statistical nature of pinning and the effect of cooling conditions make distributions of trapped vortices very cavity-dependent. 

The answer to the second part of the question depends on the rf field range. At low fields $H\ll H_c$, flux losses can be mitigated by materials defects which pin vortices at the surface, and $S$ can be further reduced by engineering a dirty layer at the surface.  At the same time, a high density of metallic pins spaced by $\lesssim \lambda$ would greatly increase the eddy current losses even without trapped vortices, whereas non-metallic pins such as nano pores would increase the BCS losses by reducing the current-carrying cross-section and increasing the rf field penetration depth. At high rf fields mitigation of flux losses by pinning becomes ineffective because it cannot provide $J_c\simeq J_d$ to counter the rf screening current density close to the depairing limit at $H\approx H_c=200$~mT.  Previous works on artificial pinning centers in high-$J_c$ superconductors \cite{jc1,jc2,jc3,jc4,jc5} have shown that the maximum $J_c\sim (0.1-0.2)J_d $ can be reached at the optimum volume fraction of pins $x_c\sim 10\%$ due to interplay of vortex pinning and current blocking by pins  \cite{agcond,anl,ag_pin}. 
Thus, even the optimum pinning structure cannot really reduce 
flux losses at $H\gtrsim 0.1H_c$, not to mention that such 
dense array of nanoprecipitate would greatly increase the eddy current and BCS losses. At high rf fields, trapped vortices can also result in a significant field dependence of $R_s(H)$, possibly contributing to the extended $Q(H)$ rise due to the Larkin-Ovchinnikov decrease of the vortex drag $\eta(v)$ with the vortex velocity \cite{lo}.  

\section{\label{sec:concl}Conclusion}

Our results of measurements of flux trapping and trapped flux sensitivity of large-grain cavities made for Nb ingots with different content of interstitial impurities suggest that the fraction of trapped flux increases with decreasing purity of the material and it is insensitive to surface treatments. On the other hand, the trapped flux sensitivity depends on the surface conditions such as the local mean free path and distribution of pinning centers.

The mean free path and frequency dependencies of the low-field trapped flux sensitivity observed on different elliptical cavities by different groups show similar correlations and universal behaviors after proper rescaling. Models of rf dissipation due to oscillating trapped vortices perpendicular to the surface can capture the behavior of flux sensitivity observed in this paper and previous works although the available data are not sufficient to determine which pinning mechanisms dominates.

Given a limited extent by which flux sensitivity at high fields can be mitigated by pinning defect nanostructure, we believe that a more efficient way of reducing vortex losses would be to optimize the cooling procedure of the cavity to minimize the amount of trapped flux.  As was shown in previous works, this can be achieved by inhomogeneous  cooling the cavity through $T_c$, which can push out a significant portion of trapped vortices \cite{HZB1,Roman1,HTC,Fnal1,HZB2,Kubo,LG,RG1} due to strong temperature gradients \cite{prb2,prb1}.

\begin{acknowledgments}
We would like to acknowledge Jefferson Lab technical staff for the cavity surface processing and cryogenic support. We thank Rongli Geng for providing the cavity G2, Ganapati Myneni and Kensei Umemori for providing the cavity KEK-R5. We also thank Mattia Checchin for providing the data from Ref.~\onlinecite{FnalF} and Matthias Liepe for providing us with the correction factor for the data from Ref.~\onlinecite{DG1}. This manuscript has been authored by Jefferson Science Associates, LLC under U.S. DOE Contract No. DE-AC05-06OR23177. The work of A. G. was supported by NSF under grant PHY 100614-010.
\end{acknowledgments}

\appendix

\section{\label{App:A}Dependencies of superconducting parameters on the mean free path}

Here we summarize dependencies of superconducting parameters on the mean free path used in our fitting of the experimental data. 
At $T\ll T_c$ the BCS theory gives an analytical formula for $\lambda$  and $\xi$ as functions of $l$ caused by scattering on nonmagnetic impurities \cite{AGsust}. Popular approximations of $\lambda(l)$ and $\xi(l)$ are:
\begin{gather}
\lambda=\lambda_0\left(1+\tilde{\xi}_0/l\right)^{1/2},
\label{lam} \\
\xi=0.74 \xi_0\left(1+\tilde{\xi}_0/l\right)^{-1/2},
\label{xi}
\end{gather}
where $\tilde{\xi}_0=0.88\xi_0$. The product $\xi\lambda$ is independent of $l$ as a consequence of the Anderson theorem, according to which the thermodynamic critical field $B_c$ is unaffected by nonmagnetic impurities. At $T\approx T_c$, this also follows from the GL result $B_c=\phi_0/2^{3/2}\lambda\xi$, whereas at $T=0$ the BCS theory gives
\begin{equation}
B_c(0)=(\mu_0N_n)^{1/2}\Delta_0,
\label{bc}
\end{equation}  
where $N_n=m^2v_F/2\pi^2\hbar^3$ is the normal density of states, and the gap $\Delta_0$ is independent of $l$. Here $\xi_0=\hbar v_F/\pi\Delta_0$ and $\lambda_0= (m/\mu_0ne^2)^{1/2}$ are the clean limit values of $\xi$ and $\lambda$ at $l\gg\xi_0$, where $v_F$ is the Fermi velocity, $m$ is the effective electron mass, and $n$ is the electron density.

The vortex drag coefficient $\eta=\phi_0^2/2\pi\xi^2\rho_n$ in the Bardeen-Stephen model is obtained assuming that the vortex core is a normal cylinder of radius $\xi$ with a bulk resistivity $\rho_n$. This implies that the mean free path is smaller than the core diameter, $l\lesssim \sqrt{l\xi_0}$, that is, the Bardeen-Stephen formula is only applicable in the dirty limit $l\ll \xi_0$. Substituting Eq. (\ref{xi}) into $\eta=\phi_0^2/2\pi\xi^2\rho_n$ and using the Drude formula for $\rho_n=p_F/ne^2 l$, gives:
\begin{equation}
\eta\simeq \frac{\pi^2\hbar n\Delta}{4E_F}\left(\frac{l}{\xi_0}+1\right),
\label{eto}
\end{equation} 
where $p_F$ and $E_F$ are the Fermi momentum and energy. Thus, $\eta$ is independent of $l$ in the dirty limit. Yet the Bardeen-Stephen model 
has been used in many works to describe moderately clean superconductors $l\gtrsim \xi_0$ for which Eq. (\ref{eto})  is not really applicable. Microscopic calculations of $\eta$ in a moderately clean limit give \cite{kopnin}
\begin{equation}
\eta\simeq \frac{\phi_0^2}{8\pi\xi_0^2\rho_n}\ln\frac{\Delta}{k_BT}.
\label{etl}
\end{equation}
Here $\eta$ exhibits a linear dependence on $l$ similar to that of Eq. (\ref{eto}). However, the use of the Bardeen-Stephen model in a moderately clean limit $l \gtrsim \xi_0$ disregards a factor $\simeq 0.25\ln(\Delta/k_BT)$ which can be essential when fitting the experimental data.

\section{Trapped flux sensitivity formulae}

Here we summarize the formulae for $R_{res}(B_0)$ obtained by solving the dynamic equation for a flexible vortex line driven by weak rf surface current and interacting with pinning centers for three characteristic configurations of trapped vortices shown in Fig. \ref{fig:pin_sketch}. In all cases, the normal state resistivity used in the calculation of $\eta$ is given by Eq.~(\ref{eq4}): $\rho_n = ( 7.48 \times 10^{-10} \mu\Omega \, \textrm{m}^2)/l$.

\subsection{\label{App:B}Pinned vortex parallel to the surface}

The minimum distance of a stable vortex segment from the surface $d_m$ is determined by the following balance of the pinning and the vortex image forces: 
 \begin{equation}
   \frac{\phi_0^2}{2\pi \mu_0 \lambda^3}K_1\left( \frac{2d_m}{\lambda} \right) = \phi_0J_c.
    \label{F_bal}
    \end{equation}
For weak pinning, $J_c\ll J_d$ the asymptotic expansion of $K_1(z)=(\pi/2z)^{1/2}e^{-z}$ at $z>1$ can be used. In this case Eq. (\ref{F_bal}) reduces to Eq. (\ref{dm}).

\subsection{Vortex perpendicular to the surface. Sparse strong pins.}

Dynamics of a perpendicular vortex segment of length $\ell$ pinned by a defect spaced by $z=\ell$ from the surface is described by the equation:
\begin{equation}
\eta\dot{u}=\varepsilon u'' +F\exp (-z/\lambda +i\omega t)
\label{deq}
\end{equation}
with the boundary condition $u(\ell,t)=0$ and $u'(0,t)=0$. Using the solution of Eq. (\ref{deq}) and the surface obtained in Ref. \cite{prb2}, the flux sensitivity $S=R_{res}/B_0$ 
can be recast to Eqs. (\ref{S_norm}) and (\ref{nu}). In the high-frequency limit, $\chi \gg 1$, Eq. (\ref{S_norm}) yields:
\begin{equation}
S_{hf} = \frac{\phi_0}{2\eta \lambda}.
\label{Shf1}
\end{equation}

\subsection{Vortex perpendicular to the surface. Weak collective pinning.}

For a vortex interacting collectively with a radom array of weak pinning centers spaced by distances smaller than $\lambda$, the dynamic equation for the vortex perpendicular to the surface takes the form:
\begin{equation}
\eta \dot{u}=\varepsilon u'' - \alpha u +F\exp (-z/\lambda +i\omega t),
\label{deqc}
\end{equation}   
where the Labusch spring constant $\alpha$ describes the averaged effect of pinning \cite{ce,blatter,lab} as discussed in Appendix C. The solution of Eq. (\ref{deqc}) which satisfies the boundary condition $u'=0$ at $z=0$ is:
\begin{equation}
u(z,t)=\frac{H_p\phi_0 e^{i\omega t}}{\alpha\lambda^2-\epsilon+i\omega\eta\lambda^2}\left( \lambda e^{-z/\lambda}-\lambda_c e^{-z/\lambda_c}\right).
\label{ul}
\end{equation}
Here the complex Campbell penetration depth $\lambda_c$ ~\cite{ce,blatter,lab} which defines the ripple length of the elastic vortex line disturbed by the rf current is given by
Eq. (\ref{camp}). The surface resistance takes the form \cite{AGsust}:
\begin{equation}
R_{res}=-\frac{2\pi B_0\mu_0\lambda^3\omega}{\phi_0 g}\mbox{Im} \left[\frac{s+2}{s(s+1)^2}\right],
\label{Rii}
\end{equation}
where $s=\lambda/\lambda_c=\sqrt{k+i\chi}$, and $k=\alpha\lambda^2/\epsilon$. 
In the high-frequency limit, $\chi \gg 1$, $S_{hf}$ is given by Eq.~(\ref{Shf1}).

To relate $R_{res}$ with the flux sensitivity in a cavity we write the quality factor in the form:
\begin{equation}
Q_0=\frac{\omega\mu_0\int H^2(\textbf{r})dV}{\int R_{res}(\textbf{r})H^2(\textbf{r})dA+R_{BCS}\int H^2(\textbf{r})dA}.
\label{Q0}
\end{equation} 
Here $R_{res}(n)=R_{res}(B_0)n({\bf r})\phi_0/B_0$ depends linearly on the density of perpendicular vortices $n({\bf r})$.  Hence,
\begin{equation}
Q_0=\frac{G}{R_{BCS}+\gamma R_{res}(B_0)},
\label{Q1}
\end{equation}
where $G$ is a geometric cavity constant and the factor $\gamma$ accounting contributions of trapped vortices at different locations on the inner cavity surface is given by Eq. (\ref{G}).

Consider a model spherical cavity with $H^2\propto\sin^2\theta$ and $n({\bf r})\phi_0/B_0=|\cos\theta|$, where $\theta$ is the 
polar angle between the direction of the dc magnetic field $\textbf{B}_0$ and the local normal unit vector to the surface. This implies a statistically uniform 
distribution of trapped vortices in the plane perpendicular to $\textbf{B}_0$ in which case:
\begin{equation}
\gamma=\int_0^{\pi}\!|\cos\theta|\sin^3\theta d\theta\left[ \int_0^{\pi}\!\sin^3\theta d\theta\right]^{-1} = \frac{3}{8}.
\label{toy}
\end{equation}
The integrand in the numerator of Eq. (\ref{toy}) is maximum at $\sin\theta=\sqrt{3}/2$ so vortices coming out of the inner cavity surface at $\theta\simeq 60^{\circ}$ 
contribute most to $S$.

\section{\label{App:C}Evaluation of the Labusch constant}
To evaluate $\alpha$ in Eq. (\ref{deqc}) we use the standard approach of the collective pinning theory \cite{blatter} for randomly-distributed weak pins, for example, small dielectric precipitates of radius $r_0 < \xi$ producing the maximum pinning energy $u_p\sim 4\pi B_c^2 r_0^3/3\mu_0$. The Larkin  pinning correlation length $L_c$ is determined by the condition that the elastic bending energy $\sim \epsilon u^2/L_c$ of a vortex segment of length $L_c$ is of the order of the pinning energy $u_p\sqrt{N}$ produced by the fluctuation number of pins $N$ within the interaction volume $r_p^2L_c$.  Here $N\sim n_p r_p^2L_c$, where $n_p$ is the volume density of pins, $r_p\sim\xi$ is a pin interaction radius, and $u\sim r_p$:
\begin{equation}
\epsilon\frac{r_p^2}{L_c}\sim (n_p L_c r_p^2)^{1/2}u_p. 
\label{cp}
\end{equation}
Hence,
\begin{equation}
L_c\sim \left(\frac{\epsilon \xi}{u_p\sqrt{n_p}}\right)^{2/3},\qquad u_p\simeq \frac{4\pi r_0^3B_c^2}{3\mu_0}.
\label{lc}
\end{equation}
Comparing Eq. (\ref{lc}) with $L_c=(\epsilon/\alpha)^{1/2}$ expressed in terms of the Labusch constant $\alpha$, yields
\begin{equation}
\alpha\sim\frac{u_p^{4/3}n_p^{2/3}}{\xi^{4/3}\epsilon^{1/3}}.
\label{al}
\end{equation}  
To see the dependence of $\alpha$ on the m.f.p., we notice that $u_p$ is independent of $l$ because of the Anderson theorem, whereas $\epsilon=\phi_0^2g/4\pi\mu_0\lambda^2=\epsilon_0(1+\tilde{\xi}_0/l)$, and $\xi\simeq \xi_0(1+\tilde{\xi}_0/l)^{-1/2}$, where $\tilde{
\xi}_0\approx 0.88\xi_0$. Thus,
\begin{equation}
\alpha=\alpha_0\left(1+\frac{\tilde{\xi}_0}{l}\right).
\label{amfp}
\end{equation}
Substituting Eq. (\ref{eps})  for $\epsilon$ and the GL formula for  $B_c=\phi_0/2^{3/2}\pi\lambda\xi$ into Eqs. (\ref{lc}) and (\ref{al}), the dimensionless pinning parameter $k_0=\alpha_0\lambda_0^2/\epsilon_0$ which we used to fit 
the experimental data can be written in the form:
\begin{equation}
k_0=\left(\frac{\lambda_0}{\ell}\right)^2\!\left(\frac{2}{3g}\right)^{4/3}\!\!\left(\frac{r_0}{\xi_0}\right)^4,
\label{k0}
\end{equation}   
where the mean pin spacing $\ell$ is defined by $n_p=\ell^{-3}$.   

The above contribution to $\alpha$ comes from $\delta T_c$ pinning caused by small precipitates of reduced (or zero) $T_c$. Another contribution to $\alpha$ comes from $\delta l$ pinning resulting from statistical fluctuations of the m.f.p. of atomic impurities. The formula for $\alpha$ is obtained in the same way as Eq. (\ref{al}) with the replacement of the density of nanoprecipiates $n_p \to n_i$ to the density of impurities $n_i$ and the elementary pinning energy at $T\approx T_c$~\cite{thun,kes,blatter}:
\begin{gather}
u_p\simeq \frac{4\pi B_c^2}{3\mu_0}r_i^3, \qquad r_i\sim (G\xi_0\sigma_0)^{1/3},
\label{up} \\
G(\xi_0/l)\approx \frac{1}{1+\tilde{\xi}_0/l}.
\label{g}
\end{gather} 
Here the effective interaction radius $r_i$ depends on  
the scattering cross-section on impurity  $\sigma_0$~\cite{thun} related to $l$ and $n_i$ by $\sigma_0n_i=l^{-1}$.
Using Eq.(\ref{al}), (\ref{up}), we obtain:
\begin{equation}
\alpha\sim\frac{u_p^{4/3}n_i^{2/3}}{\xi^{4/3}\epsilon^{1/3}}.
\label{ai}
\end{equation}
From Eqs. (\ref{up})-(\ref{g}) and $n_i=1/\sigma_0 l$, it follows that 
\begin{gather}
\alpha_i=\frac{\alpha_{i0}(\tilde{\xi}_0/l)^{2/3}}{(1+\tilde{\xi}_0/l)^{1/3}},
\label{aii} \\
\alpha_{i0}\simeq\left(\frac{4\pi B_c^2}{3\mu_0}\right)^{4/3}\!\!\!\frac{\sigma_0^{2/3}}{\xi_0^{2/3}\epsilon_0^{1/3}}.
\label{ai0}
\end{gather}
Equations (\ref{amfp}) and (\ref{aii}) show that $\delta T_c$ and $\delta l$ pinning result in different dependencies of $\alpha_i$ on the m.f.p. Here $\delta l$ pinning 
becomes ineffective in the clean limit $l\gg \xi_0$ and gives a weaker dependence of $\alpha\propto l^{-1/3}$ on $l$ than $\alpha\propto l^{-1}$ for $\delta T_c$ pinning in the dirty limit. Yet $\alpha_i$ can exceed $\alpha$ in the dirty limit if
\begin{equation}
\frac{\sigma_0^2n_i}{(\xi_0^{-1}+l^{-1})^2}\,\gtrsim\, r_0^6n_p.
\label{ineq}
\end{equation} 
Here the impurity scattering length $\sim \sqrt{\sigma_0}$ is of the order of atomic size, so that $\sqrt{\sigma_0}\ll r_0< \xi_0$, but the volume density of impurities $n_i$ can be much larger than the volume density of nanoprecipitates, $n_p\ll n_i$.



\begin{thebibliography}{30}
\bibitem{b}
The applied field $B_n$ is the same as the trapped flux $B_0$ in the theoretical formulae in the definition of $S$.
\end{thebibliography}
\end{document}